\documentclass[10pt]{iopart}
\usepackage{graphicx,epsfig,iopams,latexsym,cite}
\newcommand{\be}{\begin{equation}}
\newcommand{\ee}{\end{equation}}
\newcommand{\ba}{\begin{eqnarray}}
\newcommand{\ea}{\end{eqnarray}}
\newcommand{\nn}{\nonumber}

\begin{document}

\title{Testing the No-Hair Theorem with Observations of Black Holes in the Electromagnetic Spectrum}

\author{Tim Johannsen}
\address{Perimeter Institute for Theoretical Physics, Waterloo, ON, N2L 2Y5, Canada}
\address{Department of Physics and Astronomy, University of Waterloo, Waterloo, ON, N2L 3G1, Canada}
\eads{\mailto{tjohannsen@pitp.ca}}

\begin{abstract}

According to the general-relativistic no-hair theorem, astrophysical black holes depend only on their masses and spins and are uniquely described by the Kerr metric. Mass and spin are the first two multipole moments of the Kerr spacetime and completely determine all other moments. The no-hair theorem can be tested by measuring potential deviations from the Kerr metric which alter such higher-order moments. In this review, I discuss tests of the no-hair theorem with current and future observations of such black holes across the electromagnetic spectrum, focusing on near-infrared observations of the supermassive black hole at the Galactic center, pulsar-timing and very-long baseline interferometric observations, as well as X-ray observations of fluorescent iron lines, thermal continuum spectra, variability, and polarization.

\end{abstract}


\section{Introduction}
\label{sec:Intro}

Although general relativity has been tested and confirmed by a variety of weak-field experiments~\cite{Will14}, general relativity still remains practically untested in the strong-field regime~\cite{Psaltis08}. In fact, to date there have been only a few strong-field tests of general relativity in the context of either neutron stars (see Ref.~\cite{DeDeo03}; \cite{Psaltis08,Antoniadisetal13,Yagi14a,Yagi14b,Zhu15}) or black holes~\cite{Psaltis07,Braneworld1,Braneworld2,BambiBarausse11,BambiEfficiency,Braneworld3,Braneworld4,Bro14,Kong14,GW1,GW2,Yunes16}.

Black holes are among the most fascinating objects in the universe surrounded by matter under extreme conditions in a regime of strong spacetime curvature. The study of these compact objects can lead to a deeper understanding of their nature, their environment, and their implications for fundamental theories such as general relativity and string theory. At the same time, the wealth and precision of observations across the electromagnetic spectrum from both ground-based and space observatories is opening up ever increasing possibilities to test fundamental properties of black holes. In addition, highly-sensitive gravitational-wave detections will complement electromagnetic observations allowing for the possibility of entirely new and unexpected discoveries~\cite{GW1}. 

Despite a large amount of circumstantial evidence, the existence of an actual event horizon, the defining characteristic feature of a black hole, still remains to be verified. To date, the presence of an event horizon in black-hole candidates has only been inferred indirectly either by the exclusion of other types of compact objects with masses below about $3M_\odot$ (see Ref.~\cite{MR06} for a review), from the lack of observations of either Type~I X-ray bursts~\cite{NGM97,NGM01,NH02,MNR04} (see the discussion in Ref.~\cite{Psaltis06}) or, in the case of the supermassive black holes at the centers of the Milky Way and the galaxy M87, from the fact that these supermassive compact objects are greatly underluminous~\cite{Broderick09b,Broderick15}. Such observations indicate the absence of a hard stellar surface which most likely identifies the compact objects as black holes. The recent first direct observation of gravitational waves detected a waveform that is consistent with the inspiral of two black holes with masses of $\sim36M_\odot$ and $\sim29M_\odot$~\cite{GW1,GW2} (but see Ref.~\cite{Cardoso16}).

According to the no-hair theorem, isolated and stationary black holes in general relativity are uniquely characterized by their masses $M$, spins $J$, and electric charges $Q$ and are described by the Kerr-Newman metric~\cite{Newman65}, which reduces to the Kerr metric~\cite{Kerr63} in the case of electrically neutral black holes (which should be the case for astrophysical black holes, because any residual electric charge would quickly neutralize)~\cite{Israel67,Israel68,Carter71,Hawking72,Carter73,Robinson75,Mazur82}. This metric, then, is the unique stationary, axisymmetric, asymptotically flat, vacuum solution of the Einstein field equations which contains an event horizon but no closed timelike curves in the exterior domain. The no-hair theorem relies on the cosmic censorship conjecture~\cite{Penrose69} as well as on the physically reasonable assumption that the exterior metric is free of closed timelike curves. See Refs.~\cite{Chrusciel12,Robinson12} for reviews. Note, however, that the mathematical status of the no-hair theorem is not without controversy, principally in relation to the assumption (in the classical proof) of analyticity; see Sec.~3.4 in Ref.~\cite{Chrusciel12} for a discussion. Other no-hair theorems exist for different extensions of general relativity; see Refs.~\cite{Chrusciel12,Bertireview} for reviews.

Consequently, general relativity predicts that all astrophysical black holes are described by the Kerr metric. Nonetheless, such black holes will not be perfectly stationary in practice nor exist in perfect vacuum, because the presence of other objects or fields such as stars, accretion disks, or dark matter could alter the Kerr nature of the black hole (see, e.g., Ref.~\cite{Semerak15}). However, under the assumption that such perturbations are so small to be practically unobservable, one can argue that astrophysical black holes are indeed described by the Kerr metric.

In Boyer-Lindquist coordinates, then, the Kerr metric $g_{\mu\nu}^{\rm K}$ has the nonzero metric elements (setting $G=c=1$)
\ba
g_{tt}^{\rm K} &=&-\left(1-\frac{2Mr}{\Sigma}\right), \nn \\
g_{t\phi}^{\rm K} &=& -\frac{2Mar\sin^2\theta}{\Sigma}, \nn \\
g_{rr}^{\rm K} &=& \frac{\Sigma}{\Delta}, \nn \\
g_{\theta \theta}^{\rm K} &=& \Sigma, \nn \\
g_{\phi \phi}^{\rm K} &=& \left(r^2+a^2+\frac{2Ma^2r\sin^2\theta}{\Sigma}\right)\sin^2\theta,
\label{eq:kerr}
\ea
where
\ba
\Delta &\equiv& r^2-2Mr+a^2, \\
\Sigma &\equiv& r^2+a^2\cos^2 \theta
\label{eq:sigma}
\ea
and where $a\equiv J/M$ is the spin parameter.

Since black-hole spacetimes are asymptotically flat vacuum solutions of the Einstein equations, they must also be axisymmetric if they are stationary~\cite{Hawking72}. This implies that they can be described by a sequence of (Geroch-Hansen) multipole moments $\{M_l,S_l\}$. Due to the uniqueness of the Kerr spacetime within general relativity, all multipole moments of order $l\geq2$ are determined only by the first two, i.e., by the mass $M\equiv M_0$ and the spin $J\equiv S_1$. This fact can be expressed mathematically with the relation~\cite{Geroch70,Hansen74}
\begin{equation}
M_{l}+{\rm i}S_{l}=M({\rm i}a)^{l}.
\label{eq:kerrmult}
\end{equation}

However, the observed dark compact objects may not be Kerr black holes at all. Instead, these dark objects might be stable stellar configurations consisting of exotic fields (e.g., boson stars~\cite{FLP87}, gravastars~\cite{MM01}, black stars~\cite{Barce08}), naked singularities (e.g., \cite{MN92}), or black holes surrounded by stable scalar fields~\cite{Herdeiro14PRL,Herdeiro16}. Alternatively, the fundamental theory of gravity may be different from general relativity in the strong-field regime, and the vacuum black-hole solution might not be described by the Kerr metric. So far, a number of black hole solutions in several theories of gravity other than general relativity have already been found; see Refs.~\cite{Babichevreview,Baraussereview,Bertireview,Herdeiro15} for reviews.

As a result, observational tests of the no-hair theorem may verify both the Kerr nature of black holes and the strong-field predictions of general relativity. Unfortunately, such tests are slightly complicated by the fact that the Kerr metric is not unique to general relativity but also the most general black hole solution in a large class of scalar-tensor theories of gravity~\cite{PsaltisKerr,Sotiriou12,Graham14} (a similar property also holds for the Kerr-Newman metric~\cite{CruzDombriz09}). Nonetheless, the fact that the no-hair theorem requires the multipole moments of a stationary black hole to be locked by expression (\ref{eq:kerrmult}) allows for it to be tested quantitatively using observations of astrophysical black holes. Since the first two multipole moments (i.e., mass and spin) already specify their entire spacetimes, a promising strategy for testing the no-hair theorem, then, is to measure (at least) three multipole moments of the spacetime of a black hole~\cite{Ryan95} (see, also, Ref.~\cite{PaperI}).

A number of different experiments in the electromagnetic spectrum aim to test the no-hair theorem in the near future. In this article, I review such tests and discuss the underlying astrophysical mechanisms. Each of these tests requires an appropriate theoretical framework. For observations carried out in the weak-field regime (i.e., at radii $r\gg r_g$, where $r_g\equiv GM/c^2$ is the gravitational radius of a black hole with mass $M$), it is usually sufficient to employ a parameterized post-Newtonian framework within which suitable corrections to Newtonian gravity in flat space can be calculated~\cite{Will93}. In the strong-field regime, however (i.e., at radii $r\sim r_g$), the parameterized post-Newtonian formalism can no longer be applied and a careful modeling of the underlying spacetime in terms of a Kerr-like metric (e.g., \cite{MN92,CH04,GB06,VH10,JPmetric,VYS11,Jmetric,CPR14,Lin16}) is required instead. Strong-field tests of general relativity with black holes have also been proposed using gravitational-wave observations of extreme mass-ratio inspirals (EMRIs) and of gravitational ringdown radiation of perturbed black holes after a merger with another object. See Refs.~\cite{GairLRR,YunesLRR} for reviews.

In Secs.~\ref{sec:stars} and \ref{sec:pulsars}, I discuss tests of no-hair theorem with observations of stars around Sgr~A$^\ast$ and with pulsar-timing observations, respectively. Sec.~\ref{sec:metrics} contains a brief review of Kerr-like metrics. In Sec.~\ref{sec:EHT}, I focus on very-long baseline interferometric observations of shadows and accretion flows. Fluorescent iron lines, thermal continuum spectra, quasiperiodic variability, and X-ray polarization are discussed in Secs.~\ref{sec:ironlines}, \ref{sec:CF}, \ref{sec:variability}, and \ref{sec:polarization}, respectively. Sec.~\ref{sec:conclusions} contains my conclusions.

\section{Near-infrared observations of stars around Sgr~A$^\ast$}
\label{sec:stars}

Observations of the Galactic center region over several decades have provided strong evidence for the presence of a supermassive black hole, Sgr~A$^\ast$, surrounded by a dense nuclear star cluster with an extent of several parsecs; see Ref.~\cite{Genzel10} for a review. These observations have led to precise measurements of the mass $M$ and distance $R_0$ of Sgr~A$^\ast$. References~\cite{Ghez08,Gillessen09,Gillessen09b,Meyer12} and Refs.~\cite{Schoedel09,Do13,Chatzopoulos15} inferred the mass and distance of Sgr~A$^\ast$ from NIR monitoring stars on orbits around Sgr~A$^\ast$, the so-called S-stars, and in the old Galactic nuclear star cluster, respectively. References~\cite{Ghez08,Meyer12} obtained the measurements $M=(4.1\pm0.4)\times10^6\,M_\odot$, $R_0=7.7\pm0.4~{\rm kpc}$, while combining the results of Refs.~\cite{Gillessen09,Chatzopoulos15} yields the measurements \mbox{$M=(4.23\pm0.14)\times10^6\,M_\odot$}, $R_0=8.33\pm0.11~{\rm kpc}$~\cite{Chatzopoulos15}. In addition, the distance of Sgr~A$^\ast$ has been obtained from parallax and proper motion measurements of masers throughout the Milky Way by Ref.~\cite{Reid14} finding $R_0=8.34\pm0.16~{\rm kpc}$. There is strong consensus among these measurements that Sgr~A$^\ast$ has a mass of $\sim4\times10^6\,M_\odot$ and a distance of $\sim8~{\rm kpc}$.

The constraints on the mass and distance from the observations of stellar orbits will be improved in the near future by continued monitoring and by the use of the second-generation instrument GRAVITY for the Very Long Telescope (VLT), which is expected to achieve astrometry with a precision of $\sim10~{\rm \mu as}$ and imaging with a $\sim4~{\rm mas}$ resolution~\cite{GRAVITY}. Reference~\cite{Zhang15} simulated the precision with which the mass, distance, and spin of Sgr~A$^\ast$ as well as the orientation of the orbits of S-stars can be determined based on a fully-relativistic ray-tracing algorithm assuming a set of 120 GRAVITY observations for such stars over two to three full orbits ($\sim45~{\rm yrs}$) with astrometric and radial velocity precisions of $10~{\rm \mu as}$ and $1~{\rm km/s}$, respectively. Reference~\cite{Zhang15} showed that the mass and distance of Sgr~A$^\ast$ can be measured very precisely, while the precision of the spin measurement depends significantly on the eccentricity of the stellar orbit. Once a 30m-class optical telescope such as the TMT~\cite{TMT} or the E-ELT~\cite{EELT} will become available, the mass and distance of Sgr~A$^\ast$ are likely to be determined with a precision of $\sim0.1\%$~\cite{Weinberg05}.

In addition to the mass and spin of Sgr~A$^\ast$, the monitoring of stellar orbits may also measure the quadrupole moment of Sgr~A$^\ast$, thus testing the no-hair theorem via the relation in Eq.~(\ref{eq:kerrmult}). In a parametrized post-Newtonian approximation, a star that is sufficiently close Sgr~A$^\ast$ experiences an acceleration ${\bf a}_{\rm star}$ which is given by the equation (see, e.g., Ref.~\cite{Will93})
\ba
{\bf a}_{\rm star} &=& -\frac{M{\bf x}}{r^3} + \frac{M{\bf x}}{r^3} \left (4 \frac{M}{r} - v^2 \right ) +4\frac{M{\dot r}}{r^2} {\bf v} - \frac{2J}{r^3} \left[ 2{\bf v}\times {\bf {\hat J}}
-3 {\dot r} {\bf n}\times {\bf {\hat J}} - \frac{ 3{\bf n}({\bf h} \cdot {\bf {\hat J}}) }{r} \right] \nn \\
&& + \frac{3}{2} \frac{Q}{r^4} \left[5{\bf n}({\bf n} \cdot {\bf {\hat J}})^2
- 2 ({\bf n} \cdot {\bf {\hat J}}){\bf {\hat J}} - {\bf n} \right],
\label{eq:EOM}
\ea
where $\bf x$ and $\bf v$ are the position and velocity of the star, ${\bf n} = {\bf x}/r$, ${\dot r} = {\bf n}\cdot {\bf v}$, ${\bf h}= {\bf x} \times {\bf v}$, ${\bf {\hat J}} = {\bf J}/|J|$, and ${\bf J}$ and $Q$ are the angular momentum vector and the quadrupole moment of Sgr~A$^\ast$, respectively. The first three terms of the acceleration in Eq.~(\ref{eq:EOM}) correspond to the Schwarzschild part of the metric at first post-Newtonian order, the next term is the frame-dragging effect induced by the spin of Sgr~A$^\ast$, and the final term is the effect of the quadrupole moment at Newtonian order. There are additional quadrupolar corrections to the acceleration in Eq.~(\ref{eq:EOM}) at first post-Newtonian order, but these will be much smaller, because they have a stronger dependence on the distance of the star from Sgr~A$^\ast$.

The corrections to the Newtonian gravitational potential of Sgr~A$^\ast$ cause the orbit of the star to precess. The Schwarzschild-type corrections in Eq.~(\ref{eq:EOM}) lead to a precession in the orbital plane of the star, while the corrections induced by the spin (commonly referred to as Lense-Thirring precession) and quadrupole moment of Sgr~A$^\ast$ cause the orbit to precess both in and out of the orbital plane of the star. Using standard orbital perturbation theory, Ref.~\cite{Will08} (see, also, Refs.~\cite{Rubilar01,Zucker06,Nucita07,Iorio11,Iorio11b}) calculated the precessions per orbit of the pericenter angle, nodal angle, and inclination, which may be observed with instruments such as GRAVITY.

\begin{figure}[ht]
\begin{center}
\psfig{figure=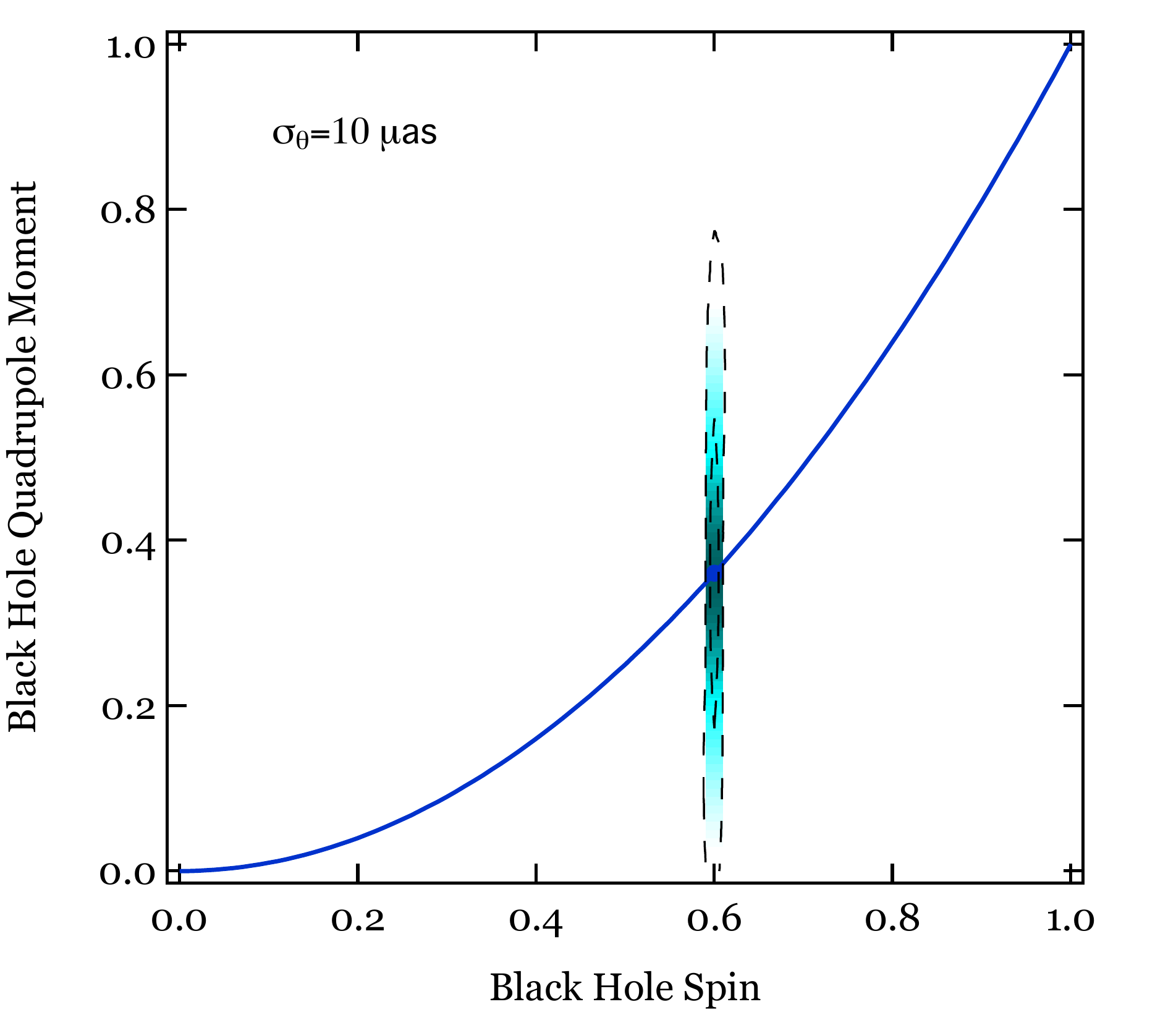,height=2.45in}
\psfig{figure=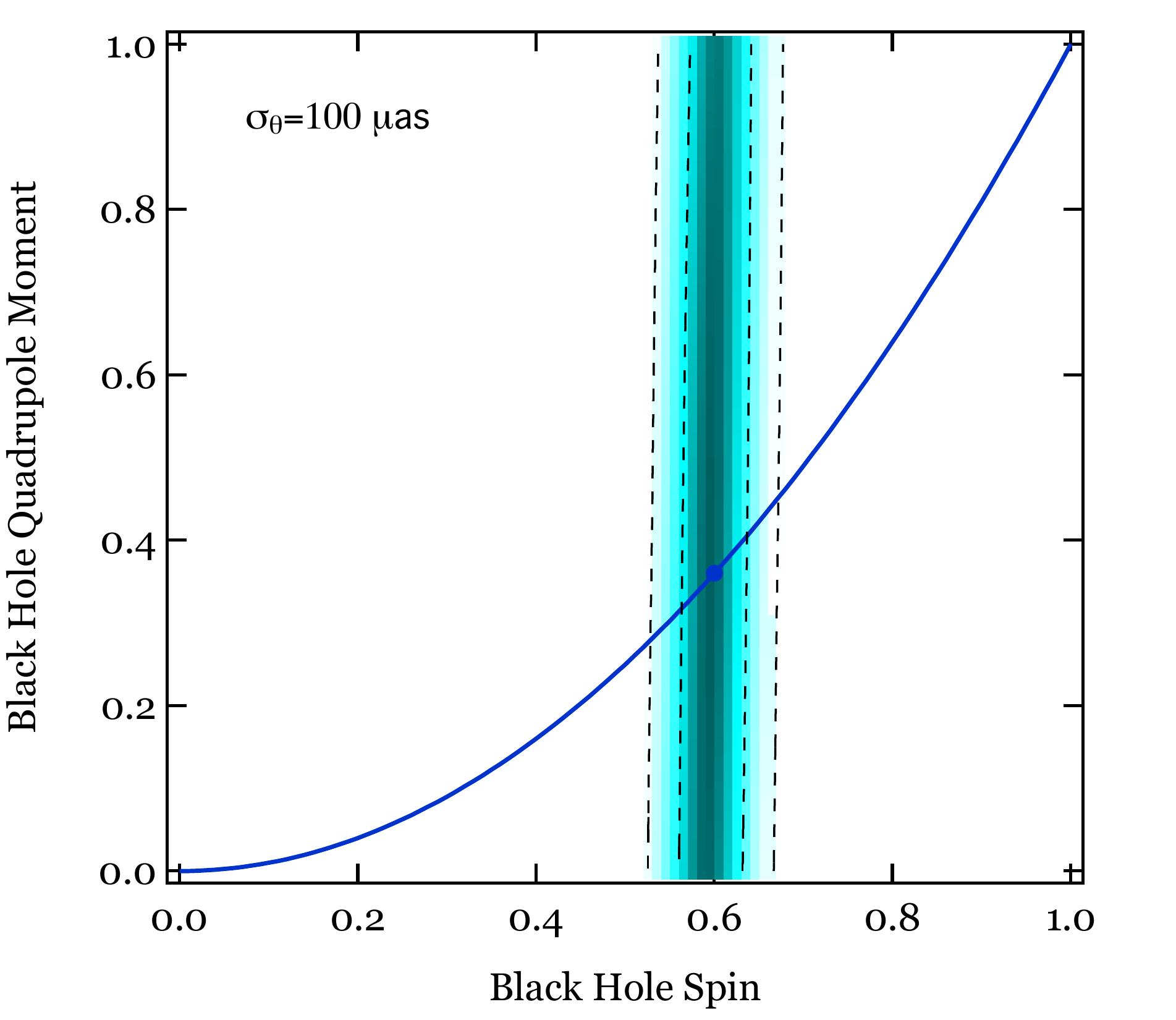,height=2.45in}
\end{center}
\caption{Posterior likelihood of measuring the spin and quadrupole moment of Sgr A* by tracing $N=40$ orbits of two stars with GRAVITY, assuming an astrometric precision of (left panel) $10~{\rm \mu as}$ and (right panel) $100~{\rm \mu as}$. The dashed curves show the 68\% and 95\% confidence limits, while the solid curve shows the expected relation between these two quantities in the Kerr metric. The filled circle marks the assumed spin and quadrupole moment of a Kerr black hole with a value of the spin $a=0.6r_g$. The two stars are assumed to have orbital separations equal to $800r_g$ and $1000r_g$ and eccentricities of 0.9 and 0.8, respectively. Even at these relatively small orbital separations, tracing the orbits of stars primarily measures the spin of the black hole, unless a very high level of astrometric precision is achieved. Taken from Ref.~\cite{PWK15}.}
\label{fig:PWKstars}
\end{figure}

Reference~\cite{PWK15} estimated the precision with which the spin and quadrupole moment of Sgr~A$^\ast$ can be measured with GRAVITY observations of the nodal and apsidal precessions of two stars with semi-major axes of $800r_g$ and $1000r_g$ and eccentricities of 0.9 and 0.8, respectively. Figure~\ref{fig:PWKstars} shows the 68\% and 95\% confidence contours of the probability density of measuring the spin and quadrupole moment of Sgr~A$^\ast$ for GRAVITY observations of such stars over $N=40$ orbits with astrometric precisions of $10~{\rm \mu as}$ and $100~{\rm \mu as}$, respectively, assuming that Sgr~A$^\ast$ is a Kerr black hole with a value of the spin $a=0.6r_g$. Even at these relatively small orbital separations, tracing the orbits of stars primarily measures the spin of the black hole, unless a very high level of astrometric precision is achieved~\cite{PWK15}.

For one S-star, Ref.~\cite{PWK15} estimated that GRAVITY observations can measure its spin with a precision
\ba
  \sigma_a &\sim&  0.064\left(\frac{\sigma_\theta}{100\,\mu{\rm as}}\right)
  \left(\frac{N}{40}\right)^{-3/2}
  \left(\frac{\tilde{a}}{1000r_g}\right)^{1/2} 
  \left(\frac{r_g/D}{5.1\,\mu{as}}\right)^{-1} \nn \\ &&
  \left[\frac{(1-e)(1-e^2)^{1/2}}{0.12}\right]
  \left(\frac{\cos i}{0.5}\right)^{-1},
\label{eq:sigmachi}
\ea
where $\sigma_\theta$ is the astrometric precision of GRAVITY observations over $N$ orbits, $D$ is the distance of Sgr~A$^\ast$, and where $\tilde{a}$, $e$, and $i$ are the semi-major axis, eccentricity, and orbital inclination of the S-star.

Although the orbits of S-stars are predominantly affected by the gravitational potential of Sgr~A$^\ast$, they may also be perturbed by other stars as well as stellar-mass black holes and, therefore, have to lie within $\sim1000$ Schwarzschild radii of Sgr~A$^\ast$ and be monitored over a sufficiently long period of time in order to measure the spin and even the quadrupole moment of Sgr~A$^\ast$~\cite{Merritt10}. Other astrophysical effects can likewise perturb the orbits of S-stars. These include: (a) the accelerations of the star due to hydrodynamic drag~\cite{Psaltisdrag12}, (b) the gravitational interaction of the star with its wake~\cite{Psaltisdrag12}, the orbital evolution of the star due to (c) stellar winds~\cite{Psaltiswinds13} and (d) tidal dissipation~\cite{Psaltiswinds13}. While the former two effects are much weaker than the effect of the quadrupole moment out to $\sim10^5$ Schwarzschild radii and, therefore, can be neglected for stars sufficiently close to Sgr~A$^\ast$~\cite{Psaltisdrag12}, the latter two effects have to be taken into account~\cite{Psaltiswinds13}.

In addition to the precession of the orbit of the star, photons emitted from the star may also be Doppler-shifted and gravitationally lensed by Sgr~A$^\ast$ and experience a corresponding time delay. These relativistic effects are strongest near the pericenter of the stellar orbit~\cite{Rubilar01,Zucker06,Angelil10,Angeliletal10,Angelil11,Angelil14}. In the next few years, already existing instruments will likely detect at least the redshift corrections due to the special-relativistic Doppler effect and the gravitational redshift~\cite{Zucker06} and the pericenter precession due to the Schwarzschild term in Eq.~(\ref{eq:EOM})~\cite{Rubilar01} for the star S2, in particular during its next pericenter passage in 2018~\cite{Ghez08,Gillessen09,Gillessen09b,Meyer12}.

Relativistic effects on the orbits of S-stars may also be imprinted on potential gravitational lensing events caused by the deflection of light rays by Sgr~A$^\ast$, which would result in the presence of two or more images of the same S-star. The position and magnification of images of gravitationally-lensed S-stars depend primarily on the mass and distance of Sgr~A$^\ast$, but may also be affected by the Schwarzschild part of the potential sourced by Sgr~A$^\ast$ or even its spin and quadrupole moment~\cite{Virbhadra98,Virbhadra99,Virbhadra02,DePaolis03,BozzaMancini04,DePaolis04,BozzaMancini05,Bozza06,Virbhadra08,BozzaMancini09,Virbhadra09,DePaolis11,BozzaGRAVITY,WeiSW15,Linet15,
Jorgensen16}. Tests of the no-hair theorem with observations of S-stars are discussed in more detail in Ref.~\cite{JohannsenReview}.

\section{Pulsar timing}
\label{sec:pulsars}

If (at least) one of the S-stars is a pulsar, timing observations of their radio pulses could provide another means to test the no-hair theorem. Observations of double pulsar systems in the Milky Way have already provided a very good testing ground for weak-field general relativity (see Refs.~\cite{Stairs03,KramerStairs08,KramerWex09} for reviews) and, in some cases, even for strong-field tests of particular theories of gravity (see Ref.~\cite{DeDeo03}; \cite{Psaltis08,Antoniadisetal13,Yagi14a,Yagi14b,Zhu15}).

Although a large number $(\sim200)$ of radio pulsars are thought to populate the stellar cluster at the Galactic center~\cite{Wharton12,Chennamangalam14}, as well as perhaps even pulsar black-hole binaries~\cite{Faucher11}, only five pulsars and one magnetar have been discovered within $15'$ of Sgr~A$^\ast$ to date, the closest of which has a distance of $\sim3''$ $(\sim1~{\rm pc})$~\cite{Morris02,Johnston06,Deneva09,Kennea13,Mori13,Rea13,Eatough13,Zadeh15}. The lack of additional pulsar detections near Sgr~A$^\ast$ is most likely caused by the strong scattering of radio waves in this region at typical observing frequencies $\sim1-2~{\rm GHz}$, requiring pulsar searches at higher frequencies~\cite{CordesLazio02}. However, pulsar detections remain elusive even though several high-frequency surveys have already been carried out near the Galactic center~\cite{Kramer00,Johnston06,Deneva09,Macquart10,Bates11,Eatough13,Siemion13} and may require highly-sensitive instruments such as the Atacama Large Millimeter/submillimeter Array (ALMA) in Chile or the SKA in Australia and South Africa~\cite{Kramer15,Keane15,Shao15}.

Nonetheless, if a pulsar sufficiently close to Sgr~A$^\ast$ is discovered, it would experience the same accelerations (and perturbations) as a regular S-star; see Eq.~(\ref{eq:EOM}). The arrival times of the emitted radio beams of the pulsar during its orbit, then, imprint characteristic time delays due to the relativistic effects, which could allow for precise measurements of the mass, spin, and quadrupole moment of Sgr~A$^\ast$. These are the Einstein time delay (caused by a combination of the relativistic Doppler effect and the gravitational redshift), the Shapiro time delay experienced by photons when passing through the gravitational potential of Sgr~A$^\ast$, and the Roemer time delay which describes the contribution of the proper motion of the pulsar to the observed time delay~\cite{Robertson38,Blandford76,DamourTaylor92,WK99,Pfahl04,Liu12,PWK15}.

Reference~\cite{Liu12} simulated the fractional precision of a mass measurement of Sgr~A$^\ast$ for a pulsar with an orbital eccentricity $e=0.5$ and inclination $i=60^\circ$ assuming weekly measurements of the pulse arrival time with an uncertainty of $100~{\rm \mu s}$ over a time span of five years. Precision levels of $10^{-6}-10^{−7}$ seem achievable~\cite{Liu12}. Similar simulations by Ref.~\cite{Liu12} showed that the spin magnitude and quadrupole moment of Sgr~A$^\ast$ can be measured with a precision of $\sim10^{-3}$ and $\sim10^{-2}$, respectively, where the quadrupole moment can be inferred with higher precision for higher values of the spin.

\begin{figure}[ht]
\begin{center}
\psfig{figure=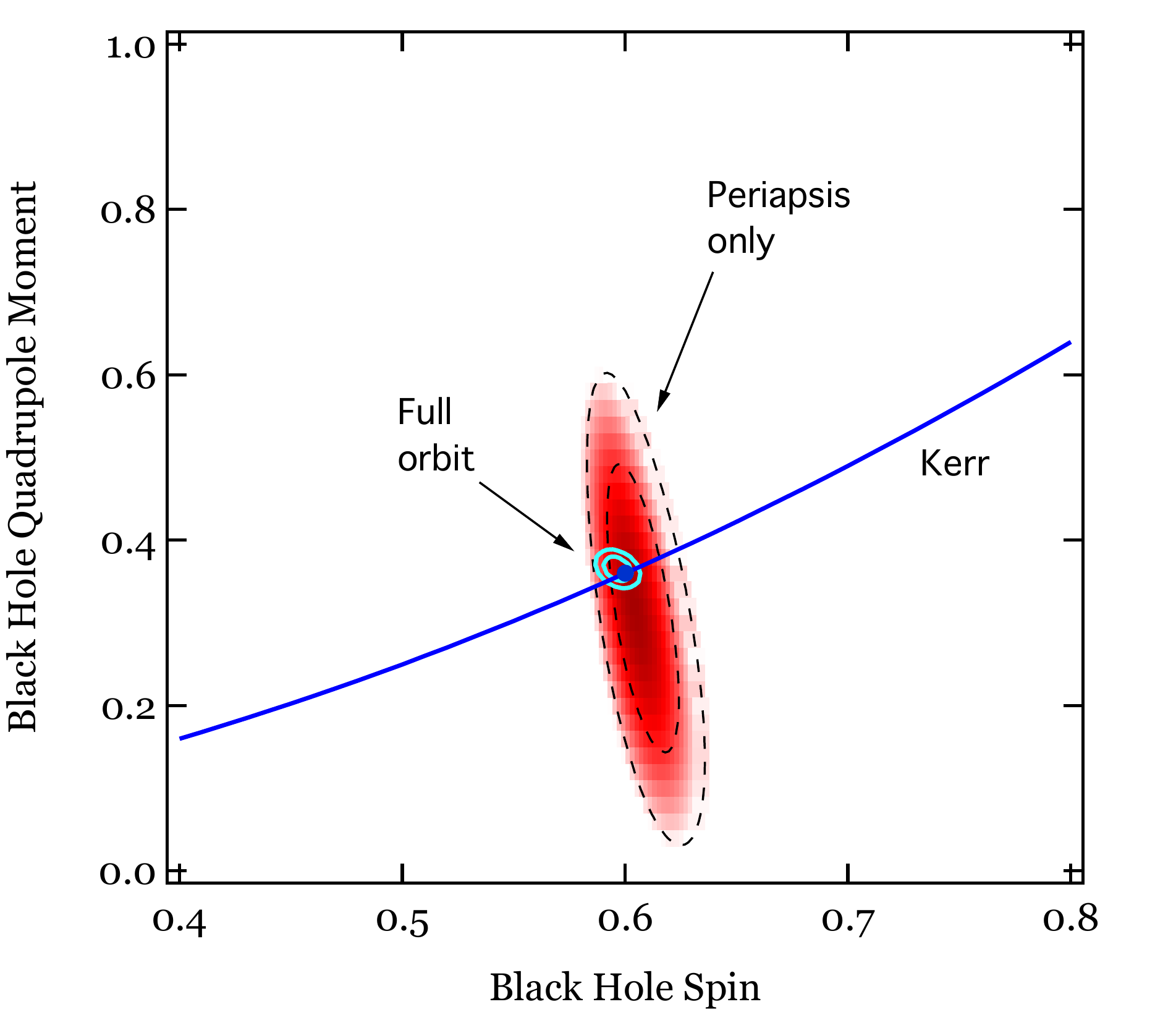,height=2.4in}
\psfig{figure=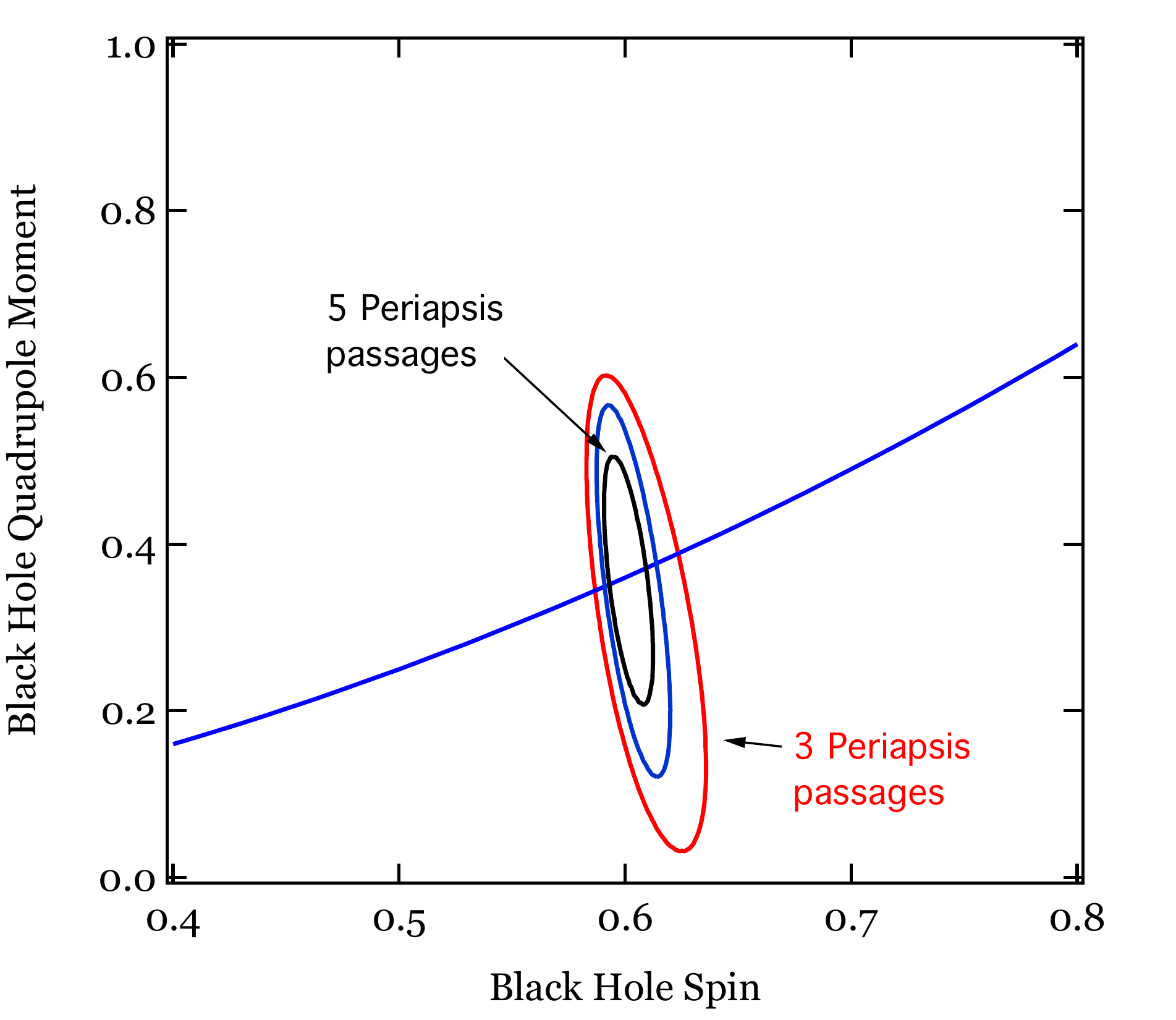,height=2.4in}
\end{center}
\caption{Simulated posterior likelihood of measuring the spin and quadrupole moment of Sgr~A$^\ast$ assuming a Kerr black hole with a value of the spin $a=0.6r_g$. In the left panel, the dashed curves show the 68\% and 95\% confidence contours, while, in the right panel, the solid curves show the 95\% confidence contours. The solid curve shows the expected relation between the spin and quadrupole moment of a Kerr black hole. The pulsar is assumed to have an orbital period of 0.5~yr (corresponding to an orbital separation of $\approx2400r_g$) and an eccentricity of 0.8, while three time-of-arrival measurements per day with equal timing uncertainty of $100~{\rm \mu s}$ have been simulated. The left panel compares the uncertainties in the measurement when only three pericenter passages have been considered in the timing solution to those when the three full orbits are taken into account. The right panel shows the increase in the precision of the measurement when the number of pericenter passages is increased from three to five. Taken from Ref.~\cite{PWK15}.}
\label{fig:PWK_Q}
\end{figure}

Reference~\cite{PWK15} refined the timing model used in Ref.~\cite{Liu12} by including higher-order post-Newtonian terms derived by Ref.~\cite{Wex95}. Figure~\ref{fig:PWK_Q} shows the posterior likelihoods of measuring the spin and quadrupole moment of Sgr~A$^\ast$ for different observing campaigns assuming a timing precision of $100~{\rm \mu s}$ and a Kerr black hole with a value of the spin $a=0.6r_g$. Even in the case of a comparably low timing precision of $100~{\rm \mu s}$ and the presence of external perturbations, a quantitative test of the no-hair theorem is possible after only a few pericenter passages and the spin and quadrupole moment of Sgr~A$^\ast$ can be measured with high precision after a few orbits~\cite{PWK15}.

Since the orbital parallax of the pulsar also makes a significant contribution to the observed timing signals, the distance of Sgr~A$^\ast$ can likewise be measured using pulsar timing. For $N$ equally distributed time-of-arrival measurements with an uncertainty $\sigma_{\rm TOA}$, the distance can be inferred with a fractional precision given by the equation~\cite{PWK15}
\ba
  \delta D &\sim& 2\,\frac{c\sigma_{\rm TOA}}{\sqrt{N}}\, \left(\frac{D}{a}\right)^2 \nn\\ 
           &\sim& 20\,{\rm pc}  \left(\frac{\sigma_{\rm TOA}}{10^2\,\mu{\rm s}}\right)  \left(\frac{N}{10^3}\right)^{-1/2}  \left(\frac{D}{8.3~{\rm kpc}}\right)^2
                  \left(\frac{a}{10^2\,{\rm au}}\right)^{-2} \;.
\ea

Reference~\cite{CPL15} calculated the Shapiro time delay experienced by photons emitted from a pulsar on an orbit around a black hole to second parameterized post-Newtonian order which also depends on the spin and quadrupole moment of the black hole (see, also, Refs.~\cite{Ashby10,Teyssandier14}). However, these effects are tiny and will primarily introduce a small bias to the measurement of the quadrupole moment discussed in Ref.~\cite{PWK15}, because the quadrupole-order time-delay and orbital effects have very different signatures on the time-of-arrival measurements~\cite{CPL15}. Reference~\cite{Pani16} showed that a binary pulsar orbiting around Sgr~A$^\ast$ could also be used as a probe of the distribution of dark matter at the Galactic center. The strong gravitational lensing of a pulsar orbiting around Sgr~A$^\ast$ could potentially also be used as a probe of certain quantum gravity effects~\cite{Pen14}. Tests of the no-hair theorem with pulsars near the Galactic center are discussed in more detail in Ref.~\cite{JohannsenReview}.

The above approach also applies to pulsars in general pulsar black-hole binaries (once discovered) and is not limited to pulsars orbiting around Sgr~A$^\ast$~\cite{WK99}. Reference~\cite{Liu14} performed mock data simulations demonstrating that a few (3--5) years of timing observations of a sufficiently compact binary harboring a pulsar and a stellar-mass black hole with future radio telescopes would allow precise measurements of the black hole mass (to $\sim0.001\%-1\%$) and spin (to $\sim1\%$). Measuring the quadrupole moment of the black hole would require extreme system configurations with compact orbits (orbital periods $\lesssim1~{\rm d}$ and eccentricities $\gtrsim0.5$) and a large ($\gtrsim10^2M_\odot$) black hole mass. Such observations can also lead to greatly improved constraints on certain alternative gravity theories even if their black hole solutions are practically identical to Kerr black holes~\cite{Liu14}.

\section{Kerr-like spacetimes}
\label{sec:metrics}

Since strong-field tests of the no-hair theorem cannot rely on the parameterized post-Newtonian formalism, a careful modeling of the underlying spacetime is required instead. Kerr-like spacetimes incorporate potential deviations from the Kerr metric via the introduction of one or more deviation parameters and reduce to the Kerr metric if all deviations vanish. Such parametrically deformed Kerr-like spacetimes typically encompasses many different theories of gravity at once and generally do not derive from the action of any particular gravity theory. The underlying theory is usually unknown and insight into this theory is hoped to be gained through observations. Several Kerr-like metrics have been proposed so far (e.g., \cite{MN92,CH04,GB06,VH10,JPmetric,VYS11,Jmetric,CPR14,Lin16}). 

If no deviations from the Kerr metric are detected by observations, the compact object is verified to be a Kerr black hole. If, on the other hand, nonzero deviations are measured, there are two possible interpretations. If general relativity still holds, the object is not a black hole but, instead, another stable stellar configuration or, perhaps, an exotic object~\cite{Hughes06}. Otherwise, the no-hair theorem would be falsified. Alternatively, within general relativity, the deviation parameters may also be interpreted as a measure of the systematic uncertainties affecting the measurement so that their effects can be treated in a quantitative manner.

Here, I focus on the Kerr-like metric of Ref.~\cite{Jmetric}. In Boyer-Lindquist-like coordinates, this metric has the nonvanishing components (setting $G=c=1$)
\ba
g_{tt} &=& -\frac{\tilde{\Sigma}[\bar{\Delta}-a^2A_2(r)^2\sin^2\theta]}{[(r^2+a^2)A_1(r)-a^2A_2(r)\sin^2\theta]^2}, \nn \\
g_{t\phi} &=& -\frac{a[(r^2+a^2)A_1(r)A_2(r)-\bar{\Delta}]\tilde{\Sigma}\sin^2\theta}{[(r^2+a^2)A_1(r)-a^2A_2(r)\sin^2\theta]^2}, \nn \\
g_{rr} &=& \frac{\tilde{\Sigma}}{\bar{\Delta} A_5(r)}, \nn \\
g_{\theta \theta} &=& \tilde{\Sigma}, \nn \\
g_{\phi \phi} &=& \frac{\tilde{\Sigma} \sin^2 \theta \left[(r^2 + a^2)^2 A_1(r)^2 - a^2 \bar{\Delta} \sin^2 \theta \right]}{[(r^2+a^2)A_1(r)-a^2A_2(r)\sin^2\theta]^2},
\label{eq:Jmetric}
\ea
where
\ba
\bar{\Delta} &\equiv& \Delta + \beta M^2,
\label{eq:beta}\\
A_1(r) &=& 1 + \sum_{n=3}^\infty \alpha_{1n} \left( \frac{M}{r} \right)^n, 
\label{eq:A1}\\
A_2(r) &=& 1 + \sum_{n=2}^\infty \alpha_{2n} \left( \frac{M}{r} \right)^n, 
\label{eq:A2}\\
A_5(r) &=& 1 + \sum_{n=2}^\infty \alpha_{5n} \left( \frac{M}{r} \right)^n, 
\label{eq:A5}\\
\tilde{\Sigma} &=& \Sigma + f(r), 
\label{eq:Sigmatilde}\\
f(r) &=& \sum_{n=3}^\infty\epsilon_n \frac{M^n}{r^{n-2}}.
\label{eq:f}
\ea

The metric of Ref.~\cite{Jmetric} contains the four free functions $f(r)$, $A_1(r)$, $A_2(r)$, and $A_5(r)$ that depend on four sets of deviation parameters (motivated by the symmetries of the Kerr metric) as well as a deviation parameter $\beta$. In the case when all deviation parameters vanish, i.e., when $f(r)=0$, $A_1(r)=A_2(r)=A_5(r)=1$, $\beta=0$, this metric reduces to the Kerr metric in Eq.~(\ref{eq:kerr}). Formally, the parametrization also includes the Kerr-Newman metric and potential deviations from it if $\beta=Q^2/M^2$, where $Q$ is the electric charge of the black hole.

The deviation functions in Eqs.~(\ref{eq:A1})--(\ref{eq:f}) are written as power series in $M/r$ (but can also be of a more general form). The lowest-order coefficients of these series vanish so that the deviations from the Kerr metric are consistent with all current weak-field tests of general relativity (c.f., Ref.~\cite{Will14}) and certain restrictions on these functions and on the deviation parameter $\beta$ exist which are determined by the properties of the event horizon. At the lowest nonvanishing order in the deviation functions, this metric depends on the parameters $\alpha_{13}$, $\alpha_{22}$, $\alpha_{52}$, $\epsilon_3$, as well as $\beta$. This metric provides a consistent description even for rapidly-spinning black holes, is free of pathological regions of spacetime on and outside of the event horizon (c.f., Ref.~\cite{pathologies}), and also admits a Carter-like constant~\cite{Jmetric}. 

Reference~\cite{Suvorov15} defined and computed multipole moments of the Kerr-Newman metric as a vacuum solution in $f(R)$ gravity theories finding that the relation of the Kerr multipole moments in Eq.~(\ref{eq:kerrmult}) is preserved in a modified form with the simple substitution $M\rightarrow\sqrt{M^2-Q^2}$ (up to a minus sign which is simply convention). Consequently, the multipole moments of the metric of Ref.~\cite{Jmetric} in the case when $\beta$ is the only nonvanishing deviation parameter are given by the relation
\begin{equation}
M_{l}+{\rm i}S_{l}=M\sqrt{1-\beta}({\rm i}a)^{l},
\label{eq:betamult}
\end{equation}
at least as long as this metric is interpreted as a vacuum solution in $f(R)$ gravity. In particular, the first three multipole moments are: $M_0=M\sqrt{1-\beta}$, $S_1=M\sqrt{1-\beta}a$, and $M_2=-M\sqrt{1-\beta}a^2$.

The metric of Ref.~\cite{Jmetric} can be mapped to known black hole solutions of specific alternative theories of gravity. These include the black hole solutions of Randall-Sundrum-type braneworld gravity (RS2;~\cite{RS2BH}), Modified Gravity (MOG;~\cite{MOG}), Einstein-dilaton-Gauss-Bonnet gravity (EdGB;~\cite{Mignemi93,Kanti96,YS11,Pani11,AY14,M15}), Chern-Simons gravity (dCS;~\cite{YP09,Pani11,YYT12}), the Bardeen metric~\cite{BardeenBH,BambiModesto13}, as well as to other Kerr-like metrics. See Ref.~\cite{JohannsenReview} for the details of these mappings and a summary of its astrophysical properties, as well as for a review of Kerr-like metrics in general.

\section{VLBI imaging of supermassive black holes}
\label{sec:EHT}

The supermassive black holes at the centers of the Milky Way, Sgr~A$^\ast$, and of the elliptical galaxy M87 (simply referred to as ``M87'' from here on) are prime targets of high-resolution very-long baseline interferometric (VLBI) observations with the Event Horizon Telescope (EHT;~\cite{Doele09a,Doele09b,Fish09}). These two sources have the largest angular sizes of known supermassive black holes in the sky (see Ref.~\cite{SMBHmasses}).

Initial VLBI observations of Sgr~A$^\ast$ in 2007--2009 at 230~GHz with a three-station array comprised by the James Clerk Maxwell Telescope (JCMT) and Sub-Millimeter Array (SMA) in Hawaii, the Submillimeter Telescope Observatory (SMTO) in Arizona, and several dishes of the Combined Array for Research in Millimeter-wave Astronomy (CARMA) in California resolved structures on scales of only $4r_S$~\cite{Doele08}, where $r_S\equiv2r_g$ is the Schwarzschild radius of Sgr~A$^\ast$. Similar observations also detected time variability on these scales in Sgr~A$^\ast$ and measured a closure phase along the Hawaii--SMA--SMTO triangle~\cite{Fish11}. In 2009--2013, follow-up observations with the same telescope array [also including the Caltech Submillimeter Observatory (CSO) in Hawaii] have led to an increased data set including numerous closure phase measurements~\cite{Fish15} and the detection of polarized emission originating from within a few Schwarzschild radii~\cite{JohnsonScience15}. 

Similar EHT observations of M87 at 230~GHz with three-station telescope arrays have also detected structure on the order of $\approx5.5$ Schwarzschild radii~\cite{DoeleM87,Akiyama15} and measured a number of closure phases~\cite{Akiyama15}. Such measurements have demonstrated the feasibility of VLBI imaging of Sgr~A$^\ast$ and M87 with the EHT on event horizon scales. 

In 2015, the existing three-station EHT array has been expanded to include ALMA in Chile, the Large Millimeter Telescope (LMT) in Mexico, the South Pole Telescope (SPT), the Plateau de Bure Interferometer (PdB) in France, and the Pico Veleta Observatory (PV) in Spain; see Ref.~\cite{FishALMA} for a recent description of the EHT. Simulations based on such enlarged telescope arrays support the possibility of probing the accretion flows of Sgr~A$^\ast$ and M87 in great detail~\cite{Doele09a,Luetal14}. The sensitivity and resolution of enlarged arrays will be greatly increased, caused primarily by ALMA which will have a sensitivity that is about 50 times greater than the sensitivity of the other stations and the long baselines from the stations in the Northern hemisphere to the SPT. In addition, this array allows for the measurement of closure phases along many different telescope triangles, some of which depend very sensitively on the parameters of Sgr~A$^\ast$~\cite{CP}, as well as of closure amplitudes along telescope quadrangles.

At around 230~GHz, the emission from Sgr~A$^\ast$ and M87 becomes optically thin allowing for an increasingly unobstructed view (see Refs.~\cite{Bro09,Luetal14} and references therein). While EHT observations of Sgr~A$^\ast$ are also affected by interstellar scattering~\cite{Bower} (which, however, become a subdominant effect above around 230~GHz~\cite{Doele08}) as well as refractive scattering~\cite{Johnson15}, EHT observations of M87 are largely unaffected by interstellar and refractive scattering. Given its much greater mass ($\sim4-6\times10^9M_\odot$~\cite{Murphy11,Walsh13}), the time scales for M87 are also longer and the rotation of the Earth is less of a challenge. In addition, the spatial scales of strong-gravity signatures are approximately comparable to those in Sgr~A$^\ast$, but the time scales for strong-gravity effects such as the orbital period of matter particles near the innermost stable circular orbit (ISCO) are much longer and, therefore, tractable via time sequenced EHT observations that allow full imaging fidelity in each epoch~\cite{DoeleM87,Akiyama15}.

The main objective of the EHT is to take the first-ever image of a black hole and to resolve its shadow~\cite{Doele09a}. Such a shadow is the projection of the circular photon orbit onto the sky along null geodesics and is expected to be surrounded by a bright ring corresponding to photon trajectories that wind around the black hole many times. Thanks to the long path length through the emitting medium of the photons that comprise the ring, these photons can make a significantly larger contribution to the observed flux than individual photons outside of the ring.

Images of shadows of Kerr black holes have been calculated by a number of different authors (e.g., Refs.~\cite{Bardeen73,Luminet79,Falcke00,Takahashi04,BeckwithDone05,PaperII,Chan13}). Ref.~\cite{Shipley16} simulated shadows of binary black holes. Images of shadows and accretion flows around non-Kerr black holes in general relativity or other theories of gravity were analyzed by, e.g., Refs.~\cite{PaperII,BambiYoshida10,Amarilla10,AE12,J13rings,Herdeiro15PRL}. Black hole shadows are also clearly visible in several (three-dimensional) general-relativistic magnetohydrodynamic simulations (GRMHD) reported to date~\cite{Moscibrodzka09,Dexter09,ShcherbakovPenna11,ChanPsaltis15a}.

The shape of the shadow of a black hole is determined only by the geometry of the underlying spacetime and, therefore, independent of the complicated structure of the accretion flow itself. For a Kerr black hole, the shape of the shadow depends uniquely on the mass, spin, and inclination of the black hole (e.g., \cite{Falcke00}). For a Schwarzschild black hole, the shadow is exactly circular and centered on the black hole. For Kerr black holes with nonzero values of the spin and the inclination, the shadow is displaced off center and retains a nearly circular shape~\cite{Takahashi04,BeckwithDone05,PaperII}, except for extremely high spin values $a\gtrsim0.9r_g$ and large inclinations, in which case the shape of the shadow becomes asymmetric~\cite{PaperII,Chan13}. However, images of black hole shadows can be significantly altered if the no-hair theorem is violated. For black holes that are described by a Kerr-like metric, the shape of the shadow can become asymmetric~\cite{PaperII} and its size can vary significantly~\cite{AE12,J13rings}. These effects (as well as a displacement of the shadow off the center of the image plane) have been quantified by Refs.~\cite{Takahashi04,PaperII,Chan13,J13rings}.

Several plausible models have been proposed for the accretion flows of Sgr~A$^\ast$ and M87, many of which are categorized as radiatively inefficient accretion flows (RIAFs; see Refs.~\cite{Falcke13,Yuan14} for reviews). Assuming that Sgr~A$^\ast$ is a Kerr black hole, Refs.~\cite{Bro09a,Bro11a} and Ref.~\cite{Bro16} inferred constraints on the spin magnitude and orientation as well as on the inclination of Sgr~A$^\ast$ based on RIAF model fits to the 2007--2009 and 2007--2013 EHT data, respectively. Ref.~\cite{Bro14} performed a similar analysis of RIAF images of Sgr~A$^\ast$ using the 2007--2009 data in the background of a quasi-Kerr metric~\cite{GB06} which contains an independent quadrupole moment. Reference~\cite{Bro14} showed that images of accretion flows in the quasi-Kerr spacetime can be significantly different from images of accretion flows around Kerr black holes revealing the asymmetric distortions of the shadow, which can be distinguished already by early EHT data. Similar images of boson stars (c.f., Ref.~\cite{Herdeiro15PRL}) at 1.3~mm were simulated by Ref.~\cite{Vincent15images}.

\begin{figure*}[ht]
\begin{center}
\psfig{figure=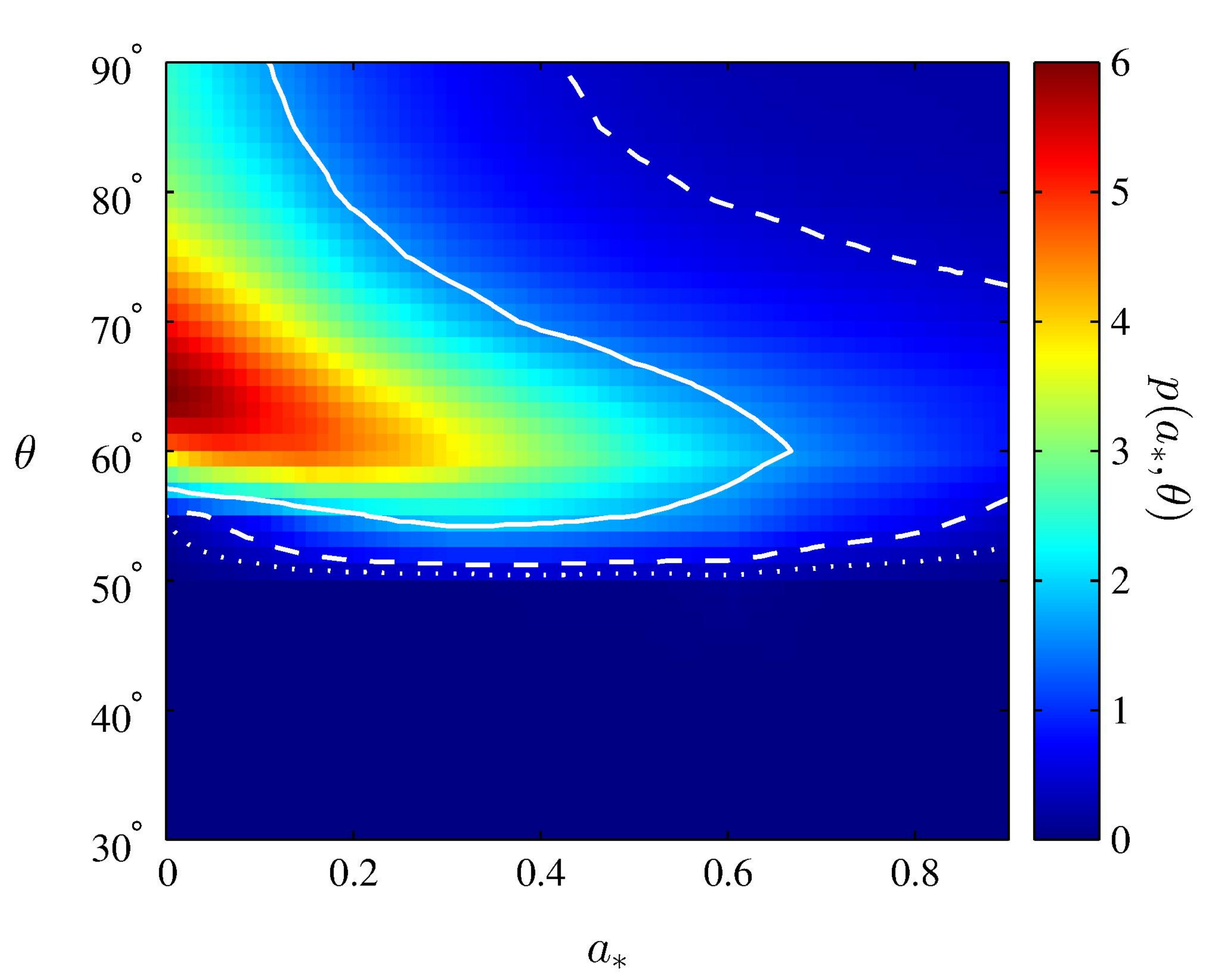,height=2.4in}
\psfig{figure=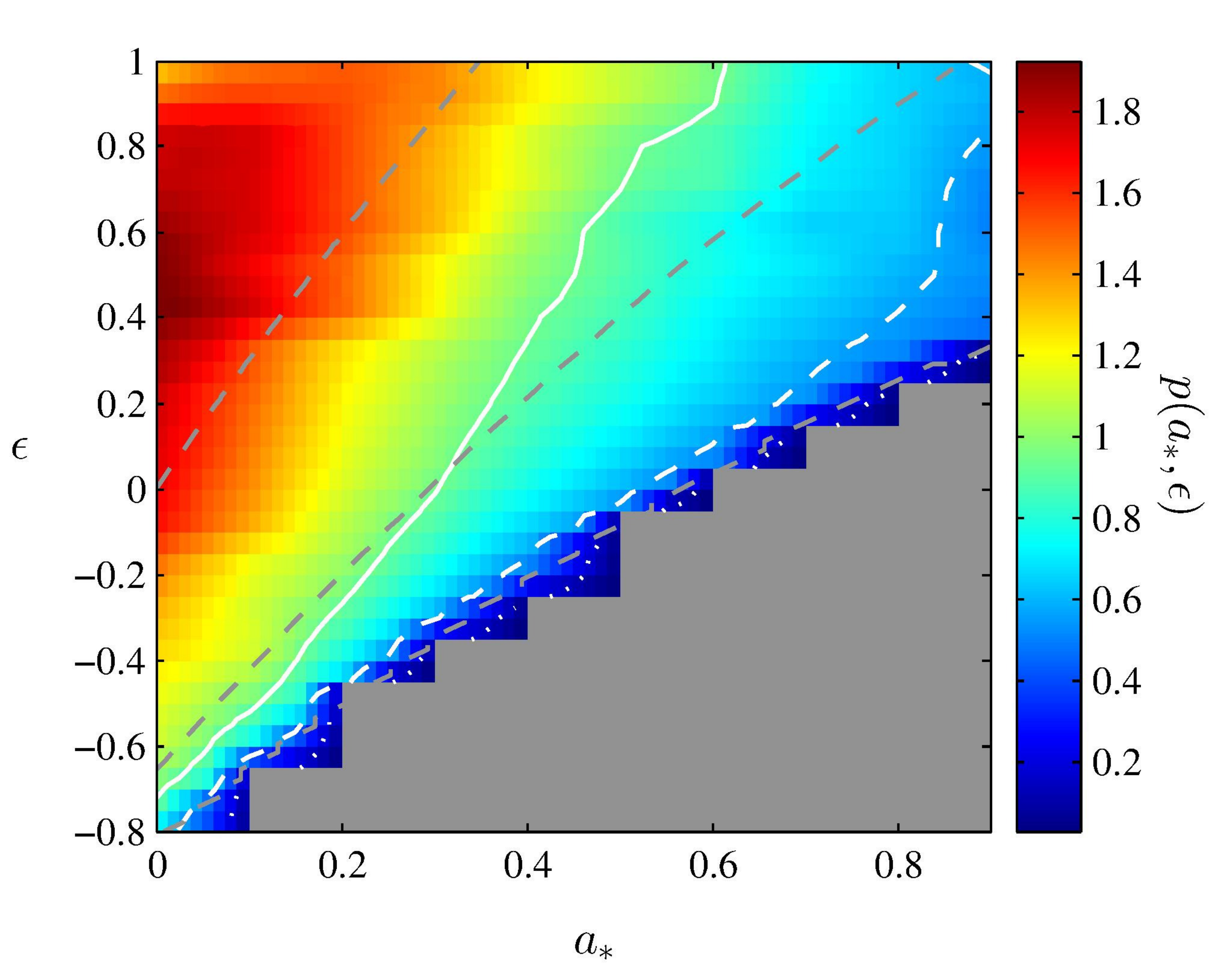,height=2.4in}
\psfig{figure=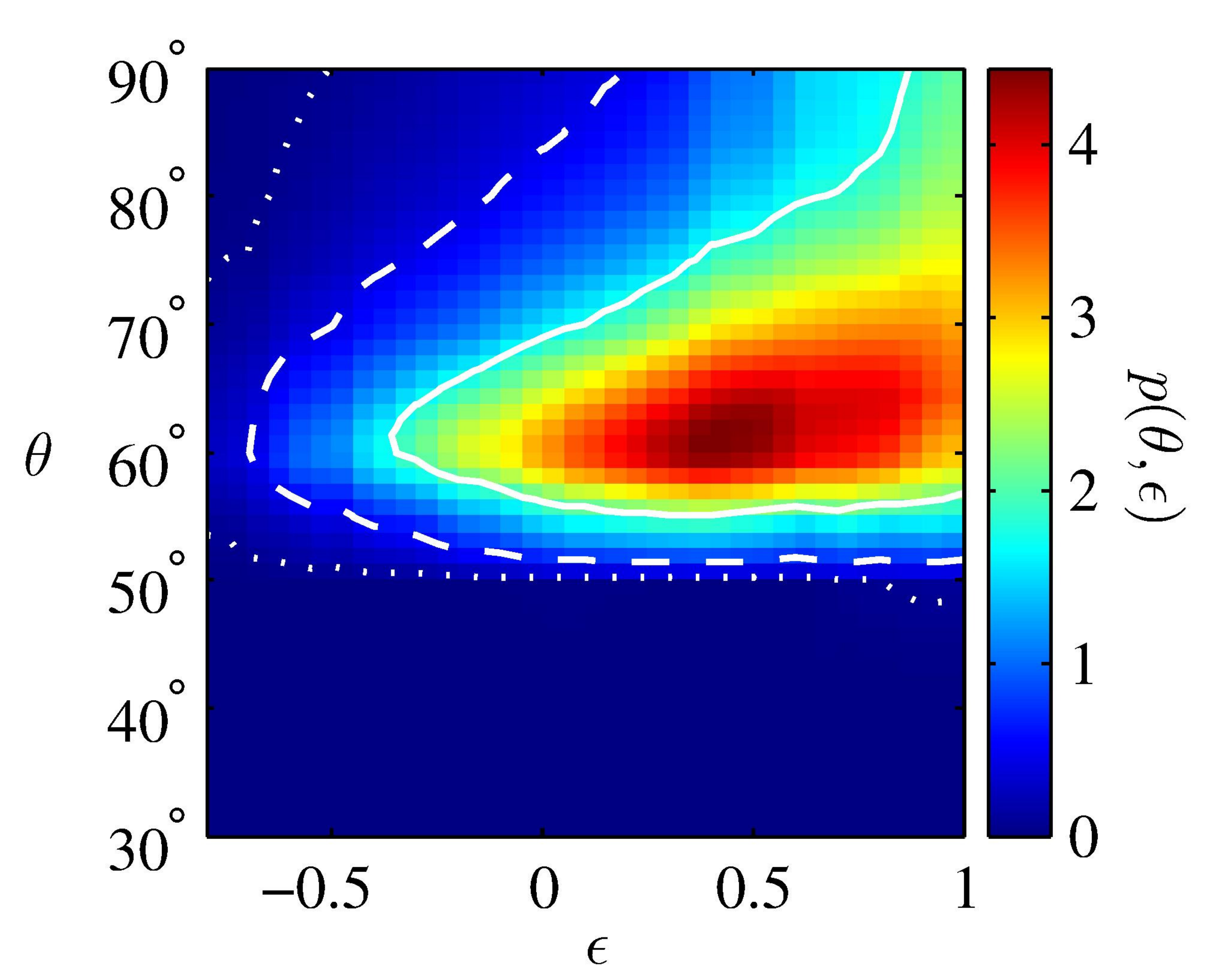,height=1.6in}
\psfig{figure=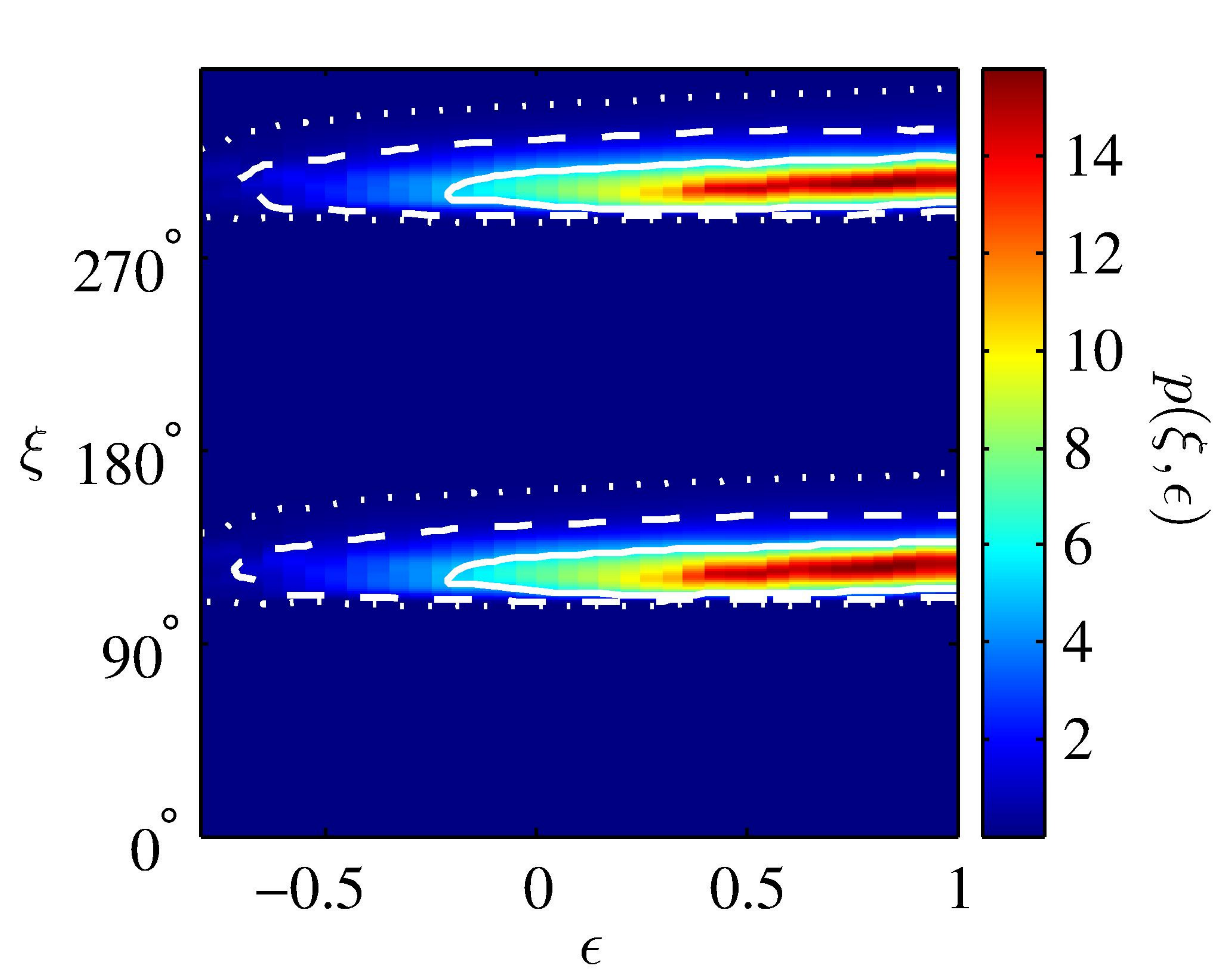,height=1.6in}
\psfig{figure=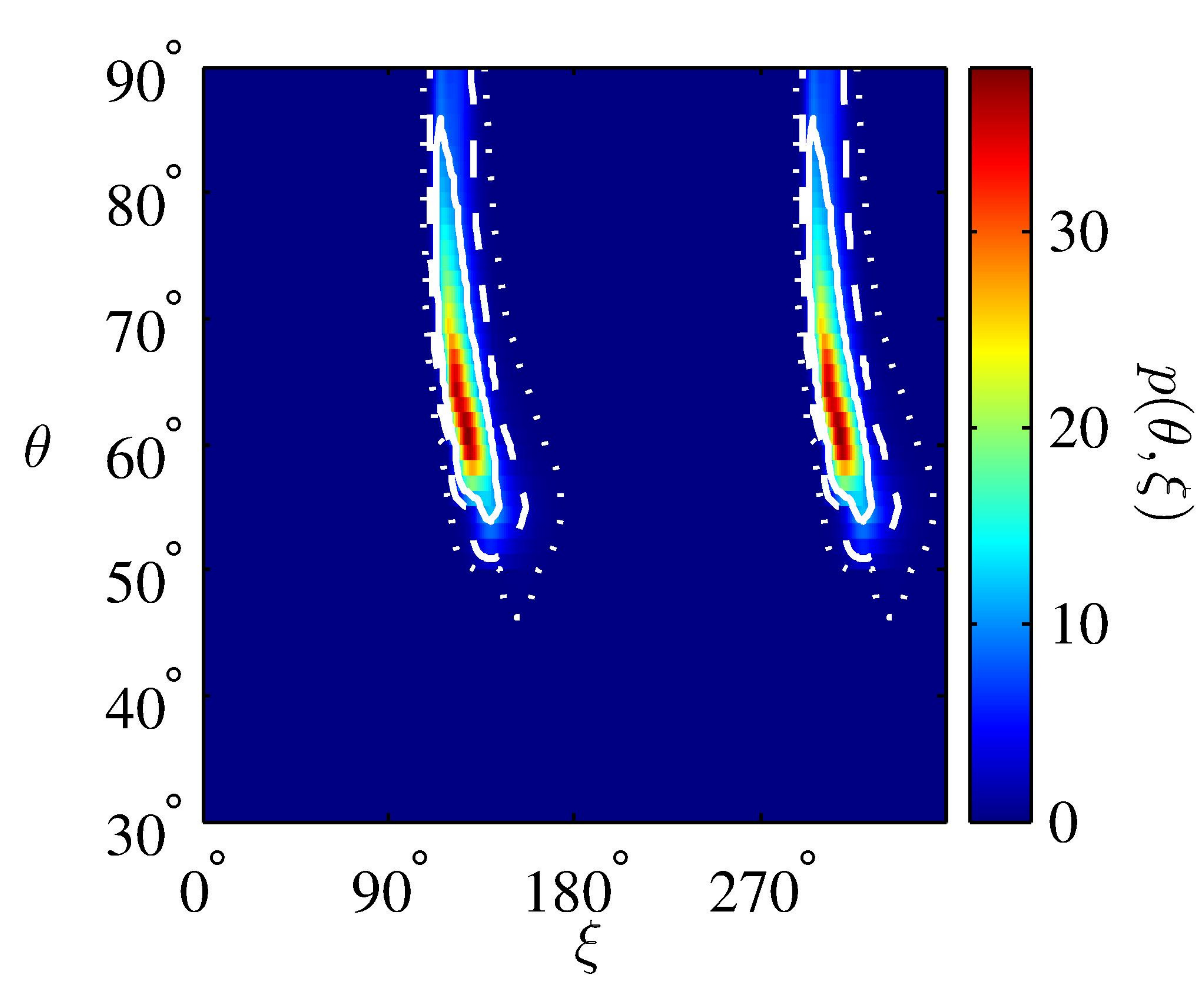,height=1.6in}
\end{center}
\caption{2D posterior probability densities as a function of (top row, left panel) dimensionless spin magnitude $a_*$ and inclination $\theta$, (top row, right panel) spin magnitude and quadrupolar deviation parameter $\epsilon$, (bottom row, left panel) inclination and quadrupolar deviation, (bottom row, center panel) spin orientation and quadrupolar deviation, and (bottom row, right panel) inclination and spin orientation, respectively marginalized over all other quantities. In each panel, the solid, dashed, and dotted lines show the $1\sigma$, $2\sigma$, and $3\sigma$ confidence regions, respectively. In the top right panel, lines of constant ISCO radius are shown as dashed gray lines, corresponding to $6r_g$, $5r_g$, and $4r_g$ from top to bottom, while the gray region in the lower right is excluded. Taken from Ref.~\cite{Bro14}.}
\label{fig:RIAF3}
\end{figure*}

Fitting the early EHT data to a library of RIAF images, Ref.~\cite{Bro14} showed that previous measurements of the inclination and spin position angle in the same RIAF model~\cite{Bro09a,Bro11a} are robust to the inclusion of a quadrupolar deviation from the Kerr metric. Figure~\ref{fig:RIAF3} shows the 2D posterior probability densities of various combinations of the spin magnitude, spin orientation, inclination, and quadrupolar deviation, each marginalized over the remaining two parameters not shown. The spin magnitude and the quadrupolar deviation are strongly correlated, roughly along lines of constant ISCO radius as shown in Fig.~\ref{fig:RIAF3}, while the spin and the inclination are only modestly correlated. The spin orientation could be determined only up to a $180^\circ$ degeneracy. Reference~\cite{Bro14} obtained constraints (with $1\sigma$ errors) on the spin magnitude $a_*=0^{+0.7}$, spin orientation $\xi=127^{\circ+17^\circ}_{~-14^\circ}$ (up to a $180^\circ$ degeneracy), and inclination $\theta=65^{\circ+21^\circ}_{~-11^\circ}$, while constraints on the deviation parameter remained weak. However, such constraints within a specific RIAF model will improve dramatically with EHT observations using larger telescope arrays~\cite{Johannsenetal16}.

Since the size of the shadow is determined primarily by the mass-distance ratio $M/D$, the existing mass and distance measurements, for which mass and distance are correlated either roughly as $M\sim D^2$ in the case of observations of stellar orbits~\cite{Ghez08,Gillessen09} or as $D\sim M^0$ in the case of the maser observations by Ref.~\cite{Reid14}, can be improved by measurements of this ratio with the EHT~\cite{SMBHmasses}. If Sgr~A$^\ast$ is indeed a Kerr black hole, then its angular radius measured by upcoming EHT observations has to coincide with the angular radius inferred from existing measurements of the mass and distance of Sgr~A$^\ast$ which constitutes a null test of general relativity~\cite{Psaltis14}.

Reference~\cite{SMBHmasses} used simple scaling arguments to estimate the precision for a measurement of the size of the shadow at a wavelength $\lambda$ with an EHT array comprised of five to six stations. Based on this estimate, Ref.~\cite{PWK15} argued that the EHT can measure the asymmetry of the shadow as defined in Reference~\cite{PaperII} with a precision of $\sigma_{\rm A}=0.9~{\rm \mu as}$. Reference~\cite{PWK15} also showed that the contours of simulated measurements of the spin and quadrupole moment of Sgr~A$^\ast$ by GRAVITY and pulsar-timing observations of S-Stars and pulsars orbiting around Sgr~A$^\ast$, respectively, are nearly orthogonal to the contours of the simulated EHT measurement and, therefore, able of reducing the uncertainty of a combined measurement significantly.

Reference~\cite{Psaltis14} estimated the accuracy with which the size of the shadow can be determined with EHT observations at 1.3~mm employing an image of a Schwarzschild black hole from GRMHD simulations of the accretion flow around Sgr~A$^\ast$~\cite{ChanPsaltis15a}. Using an edge detection scheme for interferometric data and a pattern matching algorithm based on the Hough/Radon transform, Ref.~\cite{Psaltis14} demonstrated that the shadow of the black hole in this image can be localized to within $\sim9\%$.

In practice, such an image will have to be reconstructed from observed EHT data which will be affected by other uncertainties such as electron scatter broadening, atmospheric fluctuations, and instrumental noise. Reference~\cite{deblurring} performed a reconstruction algorithm for a simulated scatter-broadened RIAF image of Ref.~\cite{Bro09a} based on a simulated one-day observing run of a seven-station EHT array assuming realistic measurement conditions. This deblurring algorithm corrects the distortions of the simulated visibilities by interstellar scattering so that the resolution of the image is predominantly determined by the instrumental beam.

\begin{figure}[ht]
\begin{center}
\psfig{figure=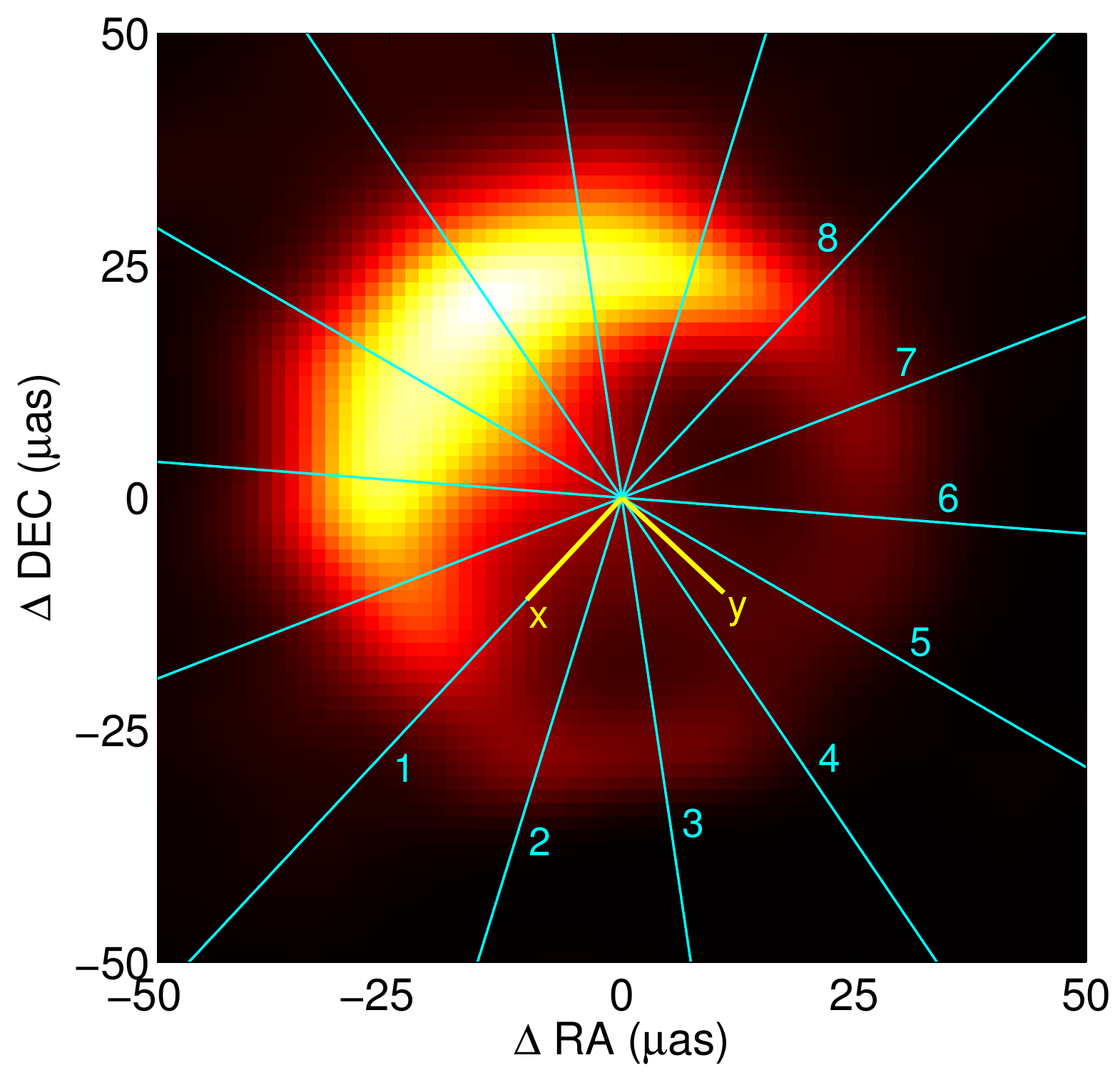,height=1.83in}
\psfig{figure=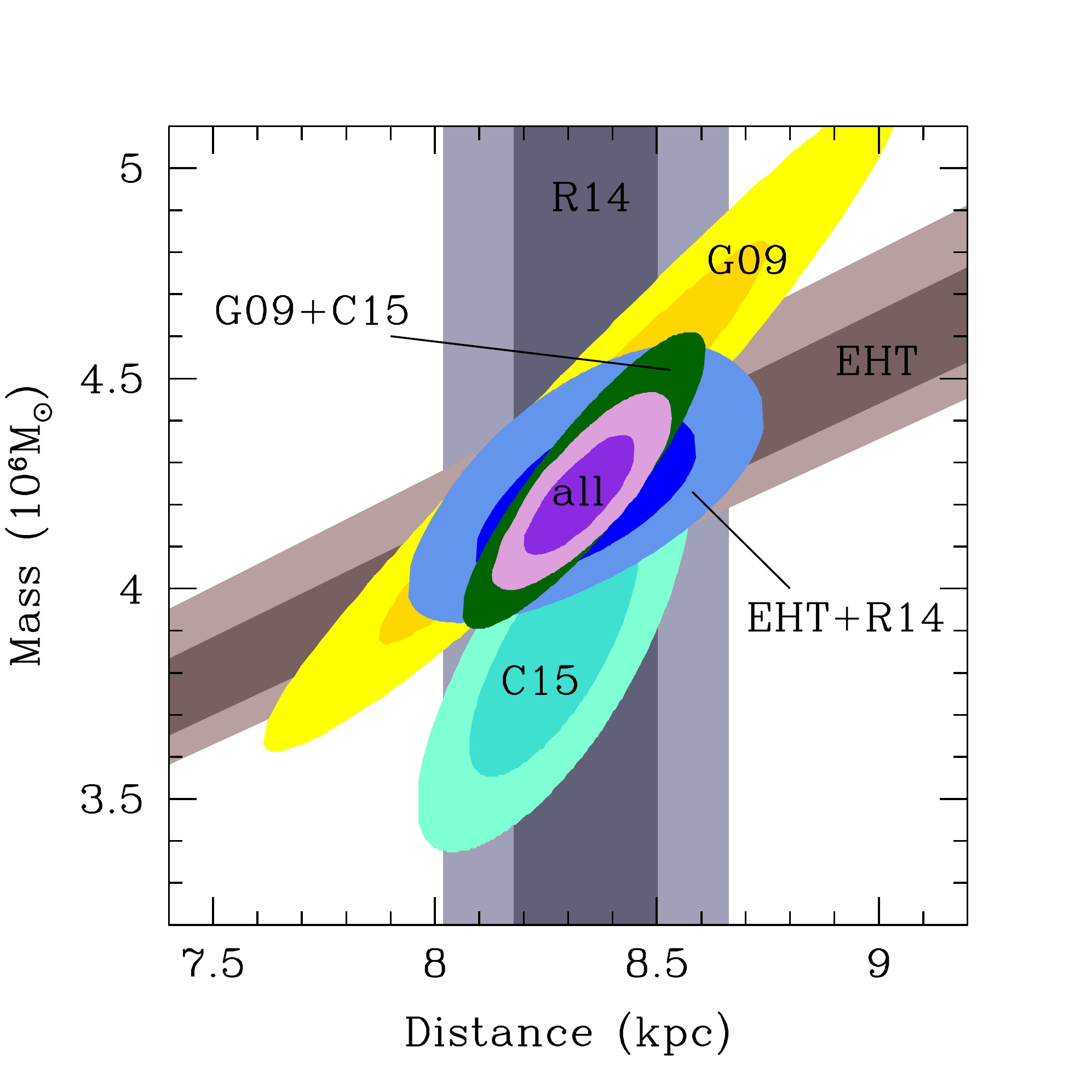,width=2.1in}
\psfig{figure=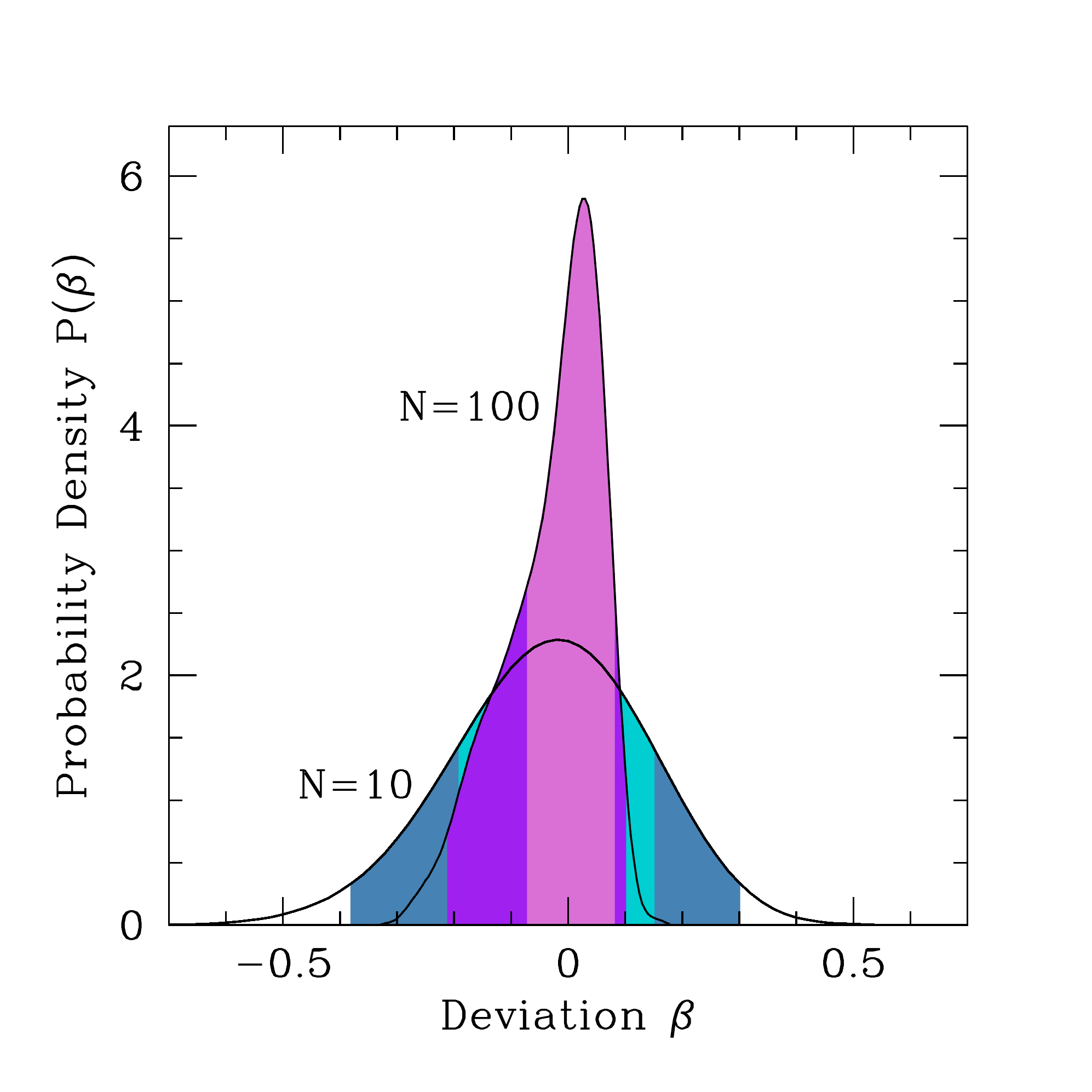,width=2.1in}
\end{center}
\caption{The left panel shows a reconstructed image of Sgr~A$^\ast$ for a simulated EHT observation at 230~GHz with a seven-station array taken from Ref.~\cite{deblurring}. The image shows seven chords for which the respective angular radii are determined from Gaussian fits of the brightness profile along the chord sections labeled ``1'',$\ldots$,``8.'' The inferred angular radius of $\approx1.5~{\rm \mu as}$ corresponds to a precision of 6\% and a length of $\approx0.16r_g$. The center panel shows $1\sigma$ and $2\sigma$ confidence contours of the probability density of the mass and distance of Sgr~A$^\ast$ for existing measurements (S-stars, ``G09''~\cite{Gillessen09}, ``R14''~\cite{Reid14}; star cluster, ``C15''~\cite{Chatzopoulos15}), a simulated measurement of the shadow size of Sgr~A$^\ast$ for $N=10$ observations with a seven-station EHT array (``EHT''), and several combinations thereof. The simulated EHT measurement improves the other constraints on the mass and distance significantly. The right panel shows simulated $1\sigma$ and $2\sigma$ confidence contours of the probability density of the deviation parameter $\beta$, corresponding to $N=10$ and $N=100$ EHT observations, each marginalized over the mass and distance using the combination of all data sets (``all'') in the $N=10$ case and of simulated stellar-orbit observations from a 30m-class telescope~\cite{Weinberg05} in the $N=100$ case. Taken from Ref.~\cite{JohannsenPRL}.
}
\label{fig:ringimages}
\end{figure}

Reference~\cite{JohannsenPRL} employed a Markov chain Monte Carlo algorithm to infer the angular radius $R$ and a potential offset $(x,y)$ from the chosen center of the shadow in this image from Gaussian fits of the brightness profile along seven chords across this image finding $R=(26.4\pm1.5)~{\rm \mu as}$, $x=(-0.3\pm1.1)~{\rm \mu as}$, $y=(1.3\pm2.2)~{\rm \mu as}$ (see the left panel of Fig.~\ref{fig:ringimages}). This estimate of the angular radius is consistent with the actual angular radius of the shadow $R\approx27.6~{\rm \mu as}$ at the $1\sigma$ level and there is no significant offset $(x,y)$ of the image center.

Reference~\cite{JohannsenPRL} combined the above simulated EHT measurement of the angular shadow radius of Sgr~A$^\ast$ with existing measurements of its mass and distance assuming a nearly circular shape of the shadow and a Gaussian distribution of the angular radius assuming a Kerr black hole with spin $a=0.5r_g$ and inclination $\theta=60^\circ$. Using the combined measurements of Refs.~\cite{Gillessen09,Chatzopoulos15,Reid14} as a prior, Ref.~\cite{JohannsenPRL} used Bayes' theorem to infer a likelihood of the mass, distance, and deviation parameters marginalized over spin and inclination. The center and right panels of Fig.~\ref{fig:ringimages} show the $1\sigma$ and $2\sigma$ confidence contours of the probability density of the mass and distance and of the deviation parameter $\beta$, respectively, for 10 EHT observations. The center and right panels of Fig.~\ref{fig:ringimages} also show the constraints on the deviation parameter $\beta$ for future measurements of the mass and distance of Sgr~A$^\ast$ obtainable with a 30m-class telescope with estimated uncertainties $\Delta M,~\Delta D\sim0.1\%$~\cite{Weinberg05} combined with simulated 100 EHT observations.

In this setup, the EHT alone can measure the mass-distance ratio (in units of $10^6\,M_\odot/{\rm kpc}$) $M/R=0.505^{+0.013+0.029}_{-0.011-0.020}$ for $N=10$ observations and $M/R=0.502^{+0.010+0.026}_{-0.005-0.007}$ for $N=100$ observations, respectively. Table~\ref{tab:massdistance} lists constraints on the mass and distance corresponding to various combinations of the EHT measurements for 10 observations with existing data showing significant improvements. In particular, combining the EHT result with the parallax measurement by Ref.~\cite{Reid14} is comparable to the mass and distance measurements from stellar orbits including the combined result of Refs.~\cite{Gillessen09,Chatzopoulos15}. If all data sets are combined as shown in the center panel of Fig.~\ref{fig:ringimages}, Ref.~\cite{JohannsenPRL} obtained the constraints on the deviation parameters $\alpha_{13}=0.1^{+0.7+1.5}_{-0.8-1.4}$, $\beta=-0.02^{+0.17+0.32}_{-0.17-0.36}$ in the $N=10$ case, while, in the $N=100$ case, they found $\alpha_{13}=-0.13^{+0.43+0.90}_{-0.21-0.34}$, $\beta=0.03^{+0.05+0.07}_{-0.10-0.24}$; the uncertainties of the mass and distance remained at the $\sim0.1\%$ level. Here, all results are quoted with $1\sigma$ and $2\sigma$ error bars, respectively.

\begin{table}[ht]
\begin{center}
\footnotesize
\begin{tabular}{lcc}
\multicolumn{3}{c}{}\\
Data   & ~~Mass ($10^6\,M_\odot$) & ~~Distance (kpc) \\
\hline
EHT+G09 & ~~$4.16^{+0.18+0.38}_{-0.16-0.31}$ & ~~$8.18^{+0.19+0.39}_{-0.19-0.37}$ \\
EHT+R14 & ~~$4.22^{+0.13+0.28}_{-0.13-0.24}$ & ~~$8.34^{+0.16+0.32}_{-0.15-0.31}$ \\
EHT+C15 & ~~$4.17^{+0.11+0.22}_{-0.11-0.21}$ & ~~$8.38^{+0.11+0.21}_{-0.11-0.21}$ \\
All     & ~~$4.22^{+0.09+0.20}_{-0.09-0.17}$ & ~~$8.33^{+0.08+0.17}_{-0.08-0.15}$ \\
\hline
\end{tabular}
\caption{Simulated mass and distance measurements using existing data (G09~\cite{Gillessen09}; R14~\cite{Reid14}; C15~\cite{Chatzopoulos15}) as priors. Taken from Ref.~\cite{JohannsenPRL}.}
\label{tab:massdistance}
\end{center}
\end{table}

The simulated constraints on the deviation parameters $\alpha_{13}$ and $\beta$ also translate into specific constraints on the parameters of known black-hole metrics in other theories of gravity; see Table~\ref{tab:devconstraints}. Note, however, that the coupling in quadratic gravity theories (i.e., theories that are quadratic in the Riemann tensor) such as EdGB has units proportional to an inverse length squared (or inverse mass squared in gravitational units). Therefore, much stronger constraints on such couplings can be obtained from observations of stellar-mass compact objects which have much lower masses and much stronger spacetime curvatures than supermassive black holes~\cite{Braneworld3,Maselli15}. While the shadow size also depends on the parameter $\alpha_{22}$, its effect is too weak to yield meaningful constraints in this scenario.

\begin{table}[h]
\begin{center}
\footnotesize
\begin{tabular}{lll}
\multicolumn{3}{c}{}\\
Theory  & ~Constraints ($N=10$) & ~Constraints ($N=100$)  \\
\hline
RS2	&  ~$\beta_{\rm tidal}=-0.02^{+0.17+0.32}_{-0.17-0.36}$  & ~$\beta_{\rm tidal}=0.03^{+0.05+0.07}_{-0.10-0.24}$  \\
MOG     &  ~$\alpha=-0.02^{+0.13+0.22}_{-0.13-0.24}$             & ~$\alpha=0.03^{+0.05+0.06}_{-0.08-0.17}$  \\
EdGB    &  ~$\zeta_{\rm EdGB}\approx0^{+0.1+0.2}_{-0.1-0.3}$     & ~$\zeta_{\rm EdGB}\approx0.022^{+0.035+0.057}_{-0.072-0.150}$ \\
Bardeen &  ~$g^2/r_g^2\approx-0.1^{+0.6+1.0}_{-0.4-0.9}$         & ~$g^2/r_g^2\approx0.09^{+0.14+0.22}_{-0.29-0.60}$ \\
\hline
\end{tabular}
\caption{Simulated $1\sigma$ and $2\sigma$ constraints on the parameters of black holes in specific theories of modified gravity (RS2~\cite{RS2BH}; MOG~\cite{MOG}; EdGB~\cite{Mignemi93,Kanti96,YS11,Pani11,AY14,M15}; Bardeen~\cite{BardeenBH,BambiModesto13}). Taken from Refs.~\cite{JohannsenPRL,JohannsenReview}.}
\label{tab:devconstraints}
\end{center}
\end{table}

At least in the case when the metric of Ref.~\cite{Jmetric} is interpreted as a vacuum solution in $f(R)$ gravity, the constraint on the parameter $\beta$ would imply a constraint on the quadrupole moment of Sgr~A$^\ast$ given by the expression $M_2=-M\sqrt{1-\beta}a^2$ [in gravitational units; see Eq.~(\ref{eq:betamult}) and Ref.~\cite{Suvorov15}]. Consequently, the above measurement of the shadow size would infer the quadrupole moment of Sgr~A$^\ast$ with a precision of $\sim9\%$ and $\sim5\%$ at the $1\sigma$ level in the $N=10$ and $N=100$ cases, respectively.

The analysis of Ref.~\cite{JohannsenPRL} estimated the shadow radius from an image of Sgr~A$^\ast$ that is constant, thus neglecting small-scale variability in the image caused by intrinsic variability of the accretion flow as well as by scatter broadening and refractive scattering along the line of sight. However, since Ref.~\cite{JohannsenPRL} fit the brightness along the chords with Gaussians, their estimate of the shadow radius is insensitive to remaining uncertainties in the interstellar scattering law. Therefore, in practice, one image of a quiescent accretion flow as the one shown in Fig.~\ref{fig:ringimages} likely corresponds to an average of several EHT observations, over which time the source variability will average out~\cite{Lu15} (but see Ref.~\cite{Medeiros16}). Likewise, the effects of different realizations of refractive substructure on different observing days will average out.

The results of Ref.~\cite{JohannsenPRL} will also be affected moderately by uncertainties in the calibration of the EHT array and in the accretion flow model used for the image reconstruction. The former imposed a $5\%$ uncertainty in early EHT observations with a three-station array, estimated from calibration for their visibility amplitudes~\cite{Fish11}. For larger telescope arrays such as the seven-station array used in this simulation, however, many more internal cross-checks will be available to improve the relative calibration of stations (the absolute calibration is not important). In particular, the use of three individual phased interferometers (Hawaii, CARMA, ALMA) that simultaneously record conventional interferometric data will permit scan-by-scan cross calibration of the amplitude scale of the array. In addition, measurements of closure phases and closure amplitudes along different telescope triangles and quadrangles are immune to calibration errors~\cite{JohannsenPRL}. See Ref.~\cite{JohannsenReview} for a review of tests of the no-hair theorem with EHT observations.

\section{Fluorescent iron lines}
\label{sec:ironlines}

Relativistically broadened iron lines are thought to originate from the X-ray emission of hot coronas around stellar-mass and supermassive black holes which irradiates off their accretion disks. Such fluorescent iron lines experience significant broadening due to the relativistic effects of light bending, Doppler boosting and beaming, and the gravitational redshift. The observed line spectra can, then, be used to measure the spin of the black hole (assuming a Kerr black hole), as well as the inclination of the accretion disk, even if the black-hole mass is unknown~\cite{Fabian89,Stella90,Laor91,Reynolds03,Dovciak04,Beckwith04,Beckwith05,Brenneman06,Reynolds08,DexterAgol09,Karas10,Dauser10}. This reflection model is strongly supported by a joint NuSTAR/Suzaku observation of IC 4329A which detected both hard X-ray emission from the corona and the corresponding reflected soft X-ray iron line emission from the accretion disk~\cite{Brenneman14}. Observations over more than a decade have resulted in numerous spin measurements of active galactic nuclei (AGN;~\cite{Brenneman06,Miniutti09,Zoghbi10,Gallo11,Brenneman11,Tan12,Lohfink12,Patrick12,Walton13,Fabian13,Reis13a,Walton14,Parker14,Keck15,Miller15,Parker15,Lohfink16}) and stellar-mass black holes~\cite{Miller09,Blum09,Reis09,Hiemstra11,Reis11,Fabian12,Steiner12b,Reis12,Miller13,Morningstar14b,King14,Garcia15,Duro16,Parker16}. See Ref.~\cite{Reynolds14} for a review, including a discussion of the underlying uncertainties of this method.

Fluorescent iron line spectra in different non-Kerr backgrounds have been analyzed by Refs.~\cite{PJ12,PaperIV,BambiIron,BambiIron2,pcGR2,Xrayprobes,Vincent14,Jiang15,Moore15,Jiang16} (see, also, Refs.~\cite{Fede15,Zhou16}). The effects of the deviation parameters on the line shapes are similar in many aspects, but a key distinction becomes evident when the metric of Ref.~\cite{Jmetric} is employed, because this metric is more general than other Kerr-like metrics in the sense that it depends on four to five independent deviation parameters (or functions) instead of just one or two (see the discussion in Refs.~\cite{Jmetric,JohannsenReview} and Sec.~\ref{sec:metrics}). Relativistically-broadened iron lines in this metric (as a function of the parameters $\alpha_{13}$, $\alpha_{22}$, $\alpha_{52}$, and $\epsilon_3$) were analyzed in Ref.~\cite{Xrayprobes} which I will discuss in the following.

In the reflection model, it is assumed that the disk plasma moves on circular equatorial orbits at the local Keplerian velocity. Furthermore, the emission from the disk is usually taken to be either isotropic or limb darkened as well as monochromatic with a rest frame energy $E_0$ (e.g., $E_0\approx6.4~{\rm keV}$ for the iron K$\alpha$ line) and  to have an emissivity profile $\epsilon(r)\propto r^{-\alpha}$, where $\alpha$ is the emissivity index that can also depend on the disk radius. The observed specific flux is, then, given by the expression
\be
F_E = \frac{1}{D^2} \int dx' \int dy' I(x',y') \delta[E_{\rm e}-E_0 g(x',y')],
\ee
where $D$ is the distance to the black hole, $I$ is the intensity, $E_{\rm e}$ and $E_0$ are the emitted and observed photon energy, respectively, $g$ is the redshift factor of a given photon, and $x'$ and $y'$ are Cartesian coordinates in the image plane.

\begin{figure}[ht]
\begin{center}
\psfig{figure=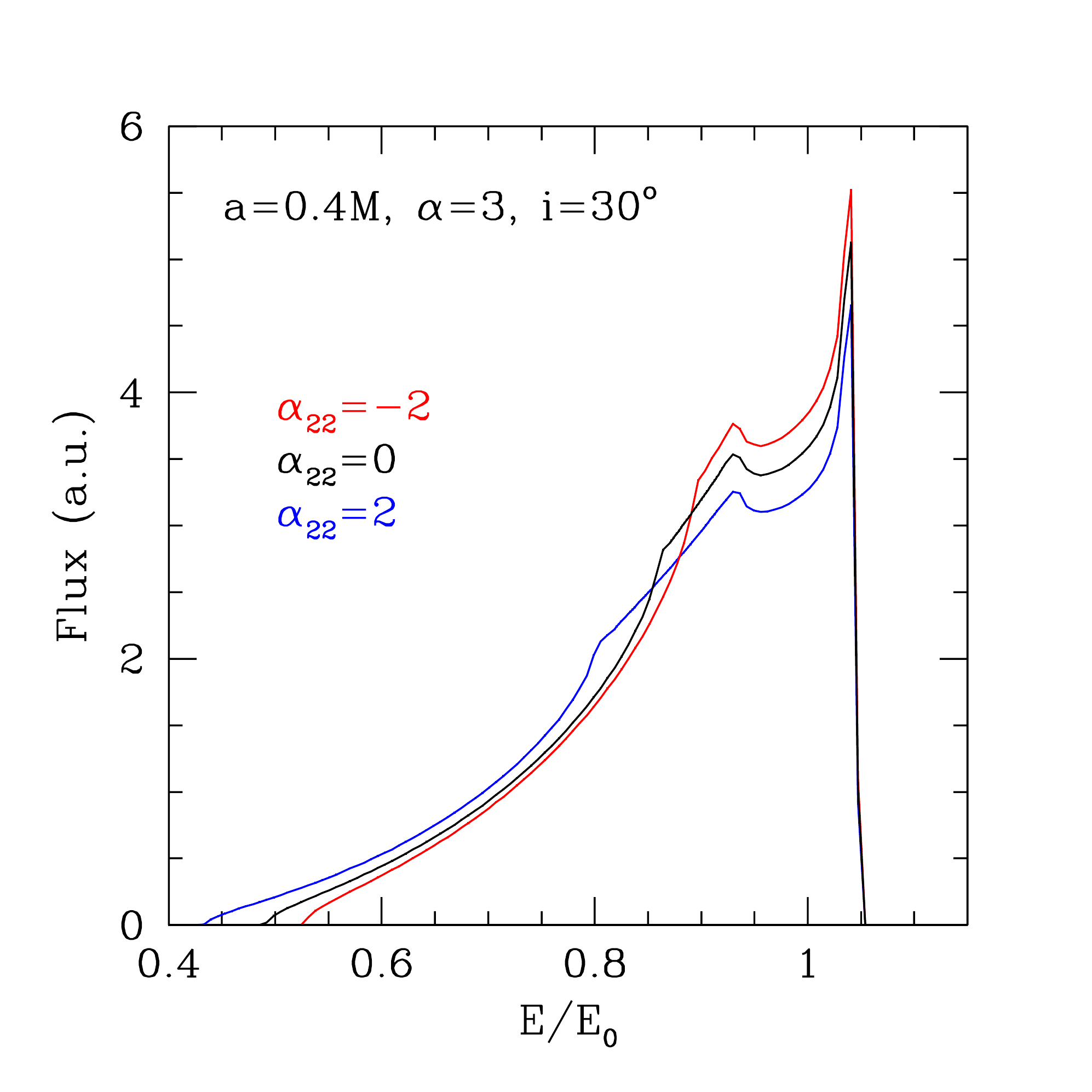,width=0.4\textwidth}
\psfig{figure=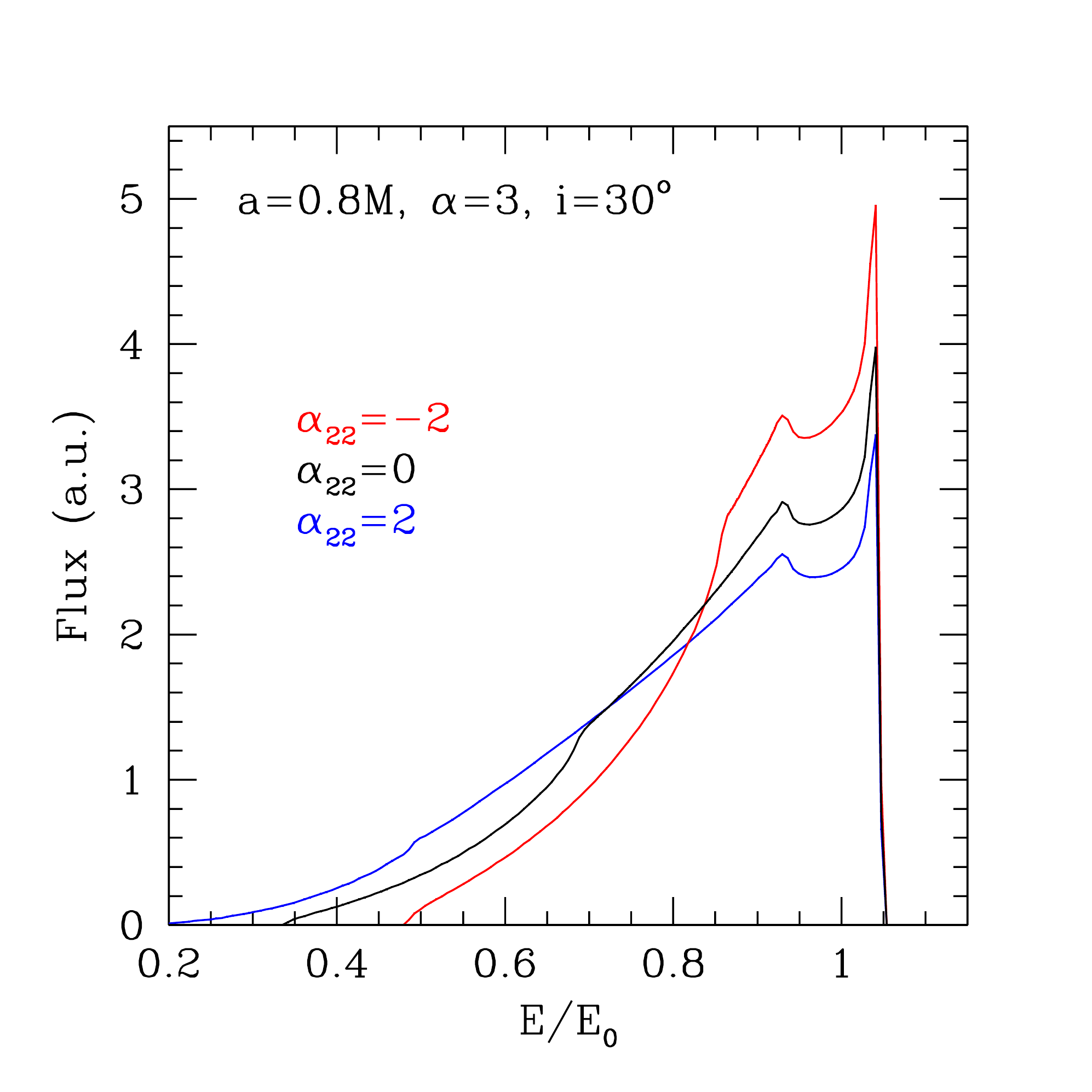,width=0.4\textwidth}
\end{center}
\caption{Iron line profiles for black holes with spins (left panel) $a=0.4r_g$ and (right panel) $a=0.8r_g$ with an outer disk radius $r_{\rm out}=100r_g$, a disk inclination $i=30^\circ$, and an emissivity index $\alpha=3$ for several values of the deviation parameter $\alpha_{22}$. The energy $E$ is measured in units of the energy of emission $E_0$. The line profiles are altered primarily at high energies and in their extent toward low energies. Taken from Ref.~\cite{Xrayprobes}.}
\label{fig:lines30}
\end{figure}

Figure~\ref{fig:lines30} shows iron line profiles for black holes with spins $a=0.4r_g$ and $a=0.8r_g$ for several values of the deviation parameter $\alpha_{22}$ for the case of isotropic emission with an emissivity index $\alpha=3$. For decreasing values of the parameter $\alpha_{22}$, the fluxes of the ``blue'' and ``red'' peaks increase and the ``red tail'' of the line profile is shortened. The first effect is primarily caused by the orbital velocity of the accretion flow, while the second effect is determined by the location of the ISCO and the photons that are emitted near the ISCO, which experience a strong gravitational redshift. The effect of the parameter $\alpha_{13}$ is very similar with the difference that the corresponding modification of the line profile as mentioned above occurs for increasing values of the parameter $\alpha_{13}$ instead of decreasing values, while the iron line profiles depend only marginally on the parameters $\epsilon_3$ and $\alpha_{52}$ if the emission is isotropic.

\begin{figure*}[ht]
\begin{center}
\psfig{figure=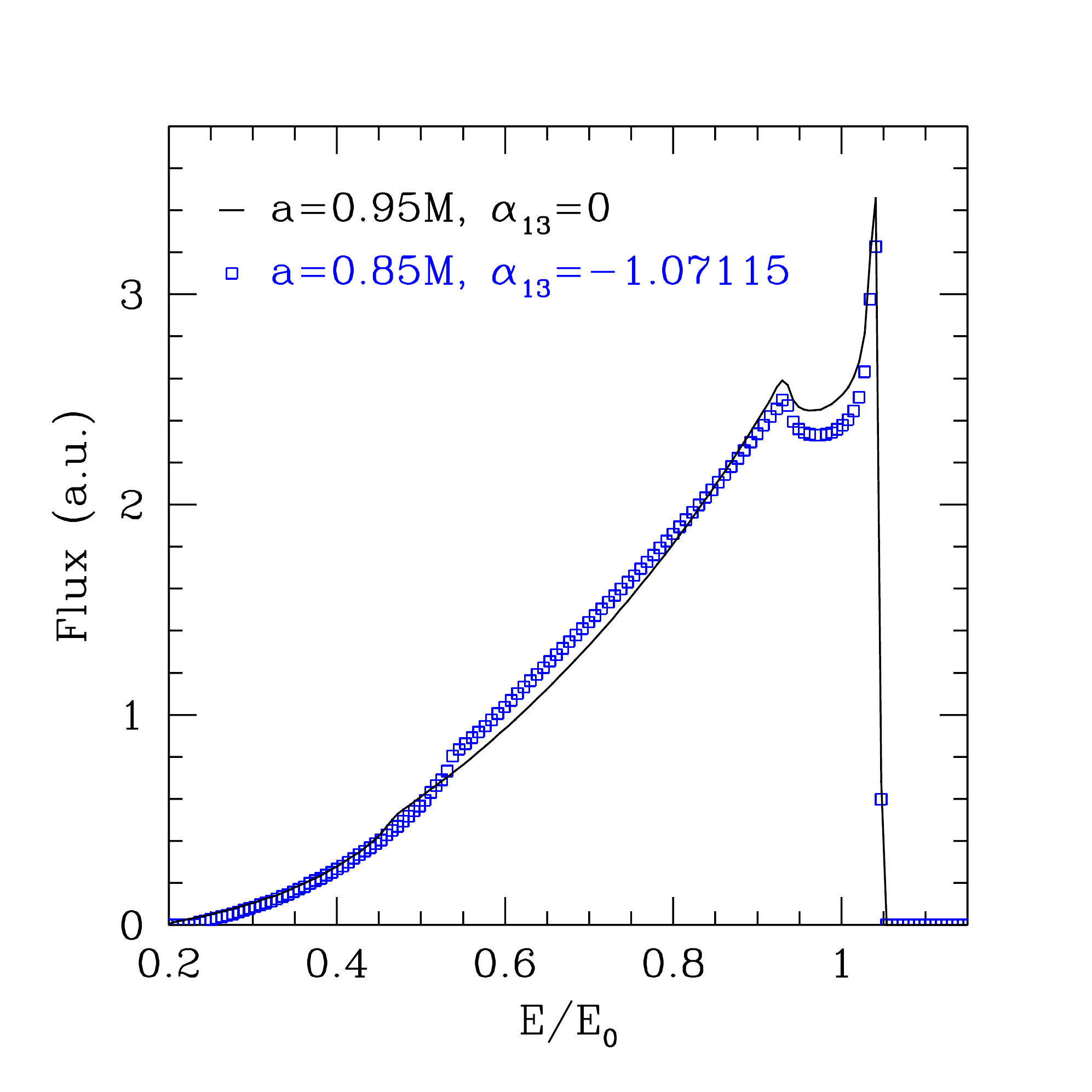,width=0.32\textwidth}
\psfig{figure=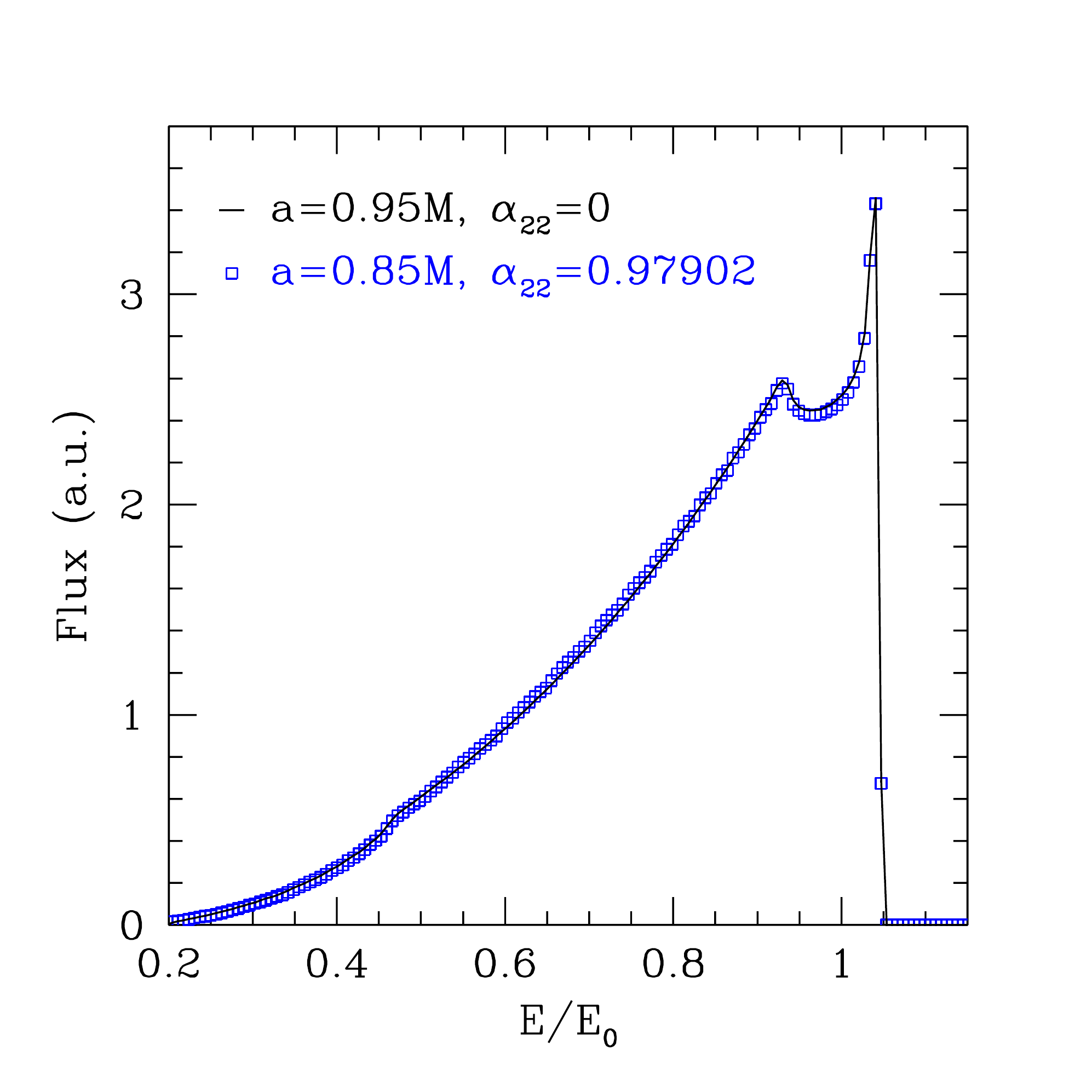,width=0.32\textwidth}
\psfig{figure=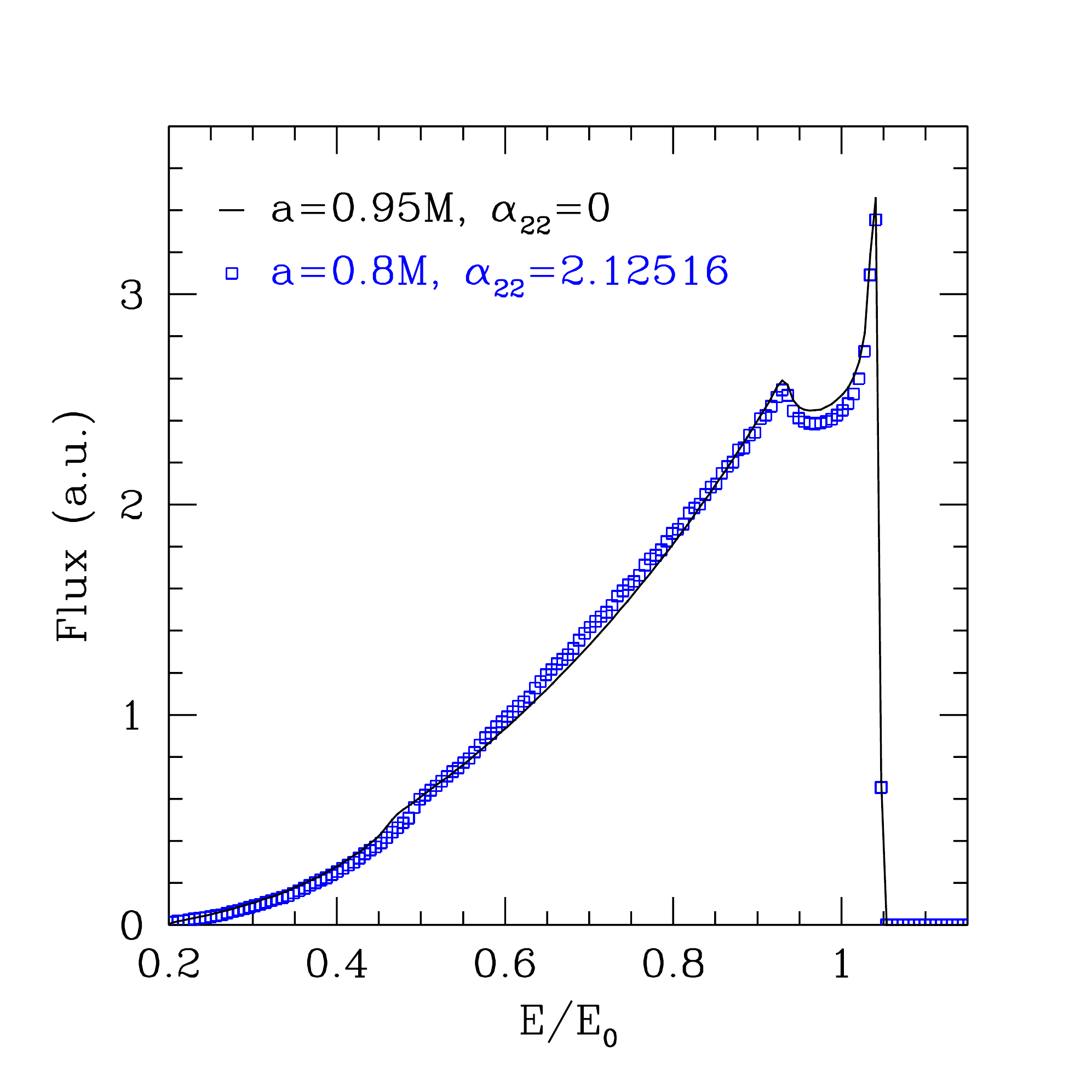,width=0.32\textwidth}
\end{center}
\caption{Iron line profiles for different values of the spin and the deviation parameters $\alpha_{13}$ (left panel) and $\alpha_{22}$ (center and right panels) such that for both sets of parameters in each panel the ISCO coincides. The other system parameters are held fixed with values $r_{\rm out}=100r_g$, $i=30^\circ$, and $\alpha=3$. For Kerr black holes with low to intermediate values of the spin, the line profiles are very similar to the profiles for Kerr-like black holes which have the same ISCO. For Kerr black holes with high spins, however, the line profiles are different from the profiles which correspond to the same ISCO radius with nonzero values of the deviation parameters $\alpha_{13}$ and $\alpha_{22}$ if the deviation is sufficiently large. Taken from Ref.~\cite{Xrayprobes}.}
\label{fig:iscolines}
\end{figure*}

At higher disk inclinations and for nonzero values of the parameters $\alpha_{13}$ or $\alpha_{22}$, the blue peak is slightly altered, while the red peak is affected only marginally. Reference~\cite{PaperIV} observed a similar effect in the case of iron line profiles described by the Kerr-like metric of Ref.~\cite{JPmetric} (which depends on only one deviation function) and identified the flux ratio of the two peaks as a potential observable of a violation of the no-hair theorem. In the metric of Ref.~\cite{Jmetric}, this effect still prevails. However, for line profiles simulated in both spacetimes (as well as the Kerr metric) the red peak is often submerged into the line profile and may be difficult to identify in practice. Similar changes of the peak flux can also be achieved by different values of the emissivity index and the outer disk radius as is well known for Kerr black holes. These parameters have to be determined from a spectral fit of the entire line profile.

If the emission is limb darkened, changing the parameters $\alpha_{13}$ and $\alpha_{22}$ has a similar effect on the line profiles as for Kerr black holes and the line profiles appear slightly narrower with an increased peak flux and, at lower inclinations, a decreased flux at lower energies. For nonzero values of the parameters $\epsilon_3$ and $\alpha_{52}$, the line profiles remain nearly unaffected at lower inclinations. At higher inclinations, however, the peak flux is modified and increases for positive values of the parameters $\epsilon_3$ and $\alpha_{52}$ and decreases for negative values of the parameters $\epsilon_3$ and $\alpha_{52}$. Note that this modification is slightly different from the one caused by nonzero values of the deviation parameters $\alpha_{13}$ and $\alpha_{22}$ for either isotropic or limb-darkened emission~\cite{Xrayprobes}.

Implementing the metric of Ref.~\cite{Jmetric} as a function of the parameter $\beta$ into the algorithm used in Ref.~\cite{Xrayprobes}, I find that the effect of this parameter on the line profiles is similar to the effect of the parameter $\alpha_{13}$ as described above. In particular, for increasing values of the parameter $\beta$, the fluxes of the blue and red peaks increase and the red tail of the line profile is shortened.

As in the case of Kerr black holes, for iron lines that originate from the Kerr-like compact objects that are described by the metric proposed by Ref.~\cite{JPmetric}, Ref.~\cite{PaperIV} found that at least for small to intermediate disk inclinations the inclination angle can be robustly determined from the location of the ``blue edge'' of the line, which depends only very little on the other system parameters including the deviation from the Kerr metric. For iron line profiles simulated in a background described by the metric Ref.~\cite{Jmetric}, this is likewise the case.

References~\cite{PJ12,PaperIV} also found that for arbitrary spins over the entire spin range iron lines of Kerr-like compact objects described by either one of the Kerr-like metrics of Refs.~\cite{GB06,JPmetric} are strongly correlated with the iron lines of Kerr black holes if the spin and deviation parameter are chosen such that the ISCO (or, more generally, the inner edge of the accretion disk~\cite{Jedges}) in both cases coincides. For nonzero values of the parameters $\alpha_{13}$ and $\alpha_{22}$ in the metric of Ref.~\cite{Jmetric}, such profiles are likewise practically indistinguishable if the Kerr black hole is spinning slowly to moderately. For Kerr black holes with high spins, however, this strong correlation between the spin and the deviation parameters $\alpha_{13}$ and $\alpha_{22}$ does not persist and the line profiles can be significantly different from each other if the deviation is large enough (\cite{Xrayprobes}; see Fig.~\ref{fig:iscolines}). This is likewise the case for sufficiently large (positive or negative) values of the parameter $\beta$.

In their initial study of relativistically broadened iron lines in a non-Kerr background, Ref.~\cite{PJ12} suggested that this strong correlation should indeed be broken for rapidly spinning black holes, because their accretion disk extends almost all the way down to the event horizon, where the effect of a deviation from the Kerr metric is most apparent. This is indeed the case for deviations described by the parameters $\alpha_{13}$, $\alpha_{22}$~\cite{Xrayprobes} and $\beta$. In contrast, in the metric of Ref.~\cite{JPmetric} (which only depends on one deviation function), the corresponding effect was not seen~\cite{PaperIV}. In the metric of Ref.~\cite{CPR14}, however, which generalizes the metric of Ref.~\cite{JPmetric} to include two deviation functions modifying the $(t,t)$ and $(r,r)$ components of the metric seperately, the strong correlation between spin and the deviations affecting the $(t,t)$ part of the metric seems to be weakened, which would explain the fact that this type of deviation can be more strongly constrained observationally in certain cases as found by Ref.~\cite{Jiang15}. Since the parameters $\alpha_{13}$, $\alpha_{22}$, and $\beta$ likewise alter the $(t,t)$ component of the metric of Ref.~\cite{Jmetric} [and, apart from the parameter $\beta$, not the $(r,r)$ component], this seems to suggest that a strong correlation regarding iron line profiles between the spin and deviations affecting the $(t,t)$ part of a given Kerr-like metric is typically reduced. Note that the $(t,\phi)$ and $(\phi,\phi)$ components of the metrics of Refs.~\cite{JPmetric,Jmetric} depend on all of these deviation parameters, respectively.

Observational constraints on deviations from the Kerr metric in other Kerr-like spacetimes based on the iron-line method have been simulated by Refs.~\cite{PaperIV,Jiang15,Avendano16b} finding that increasingly tight constraints can be obtained for black holes with ISCO radii located at $\approx r_g$ (see, also, Ref.~\cite{BambiReview}). Such constraints will be similar in magnitude for the parameters $\alpha_{13}$, $\alpha_{22}$, and $\beta$ (c.f., the discussion in Ref.~\cite{Xrayprobes}). A full analysis of original iron-line data in a non-Kerr background still remains to be performed.

In addition to the (time-averaged) spectra discussed above, fluorescent iron lines can also be observed in the time domain through X-ray reverberation and time lags~\cite{Reynolds99,Armitage03,Wilkins13,Cackett14}. Such time-dependent iron line profiles in a Kerr-like metric were discussed by Refs.~\cite{Jiang15,Jiang16,Krawcz16,Avendano16}. Reference~\cite{Kinch16} computed (time-averaged) iron line profiles based on a GRMHD simulation of an accretion flow~\cite{Noble09} around a Schwarzschild black hole.

\section{Thermal continuum spectra}
\label{sec:CF}

X-ray spectra of stellar-mass black holes often exhibit a strong thermal component, particularly in the high/soft state. Given the fact that the masses of many stellar-mass black holes have been measured quite accurately (see, e.g., Ref.~\cite{MR06}), such thermal continuum spectra can be used to measure the spin of these black holes via the so-called continuum fitting method if their distances and inclinations are known~\cite{Li05}. This method is based on a measurement of the location of the ISCO, which is remarkably constant over the course of many observations with different X-ray missions~\cite{Steiner10}. To date, the spins of eleven stellar-mass black holes have been measured in this manner~\cite{Shafee06,Davis06,McClintock06,Liu08,Gou09,Gou10,Gou11,Steiner11,Steiner12,Steiner14,Gou14,Morningstar14,Morningstar14b,Middleton14,Chen16}. See Ref.~\cite{McCl14} for a review, including a discussion of the systematic uncertainties of this method.

The continuum fitting method assumes that a given black hole is surrounded by a standard relativistic thin (Novikov-Thorne) accretion disk~\cite{Nov73} which lies in the equatorial plane of the black hole and extends from the ISCO to some outer radius $r_{\rm out}$ (typically $r_{\rm out}\sim10^5-10^6r_g$). Further, this method assumes that the disk particles move on nearly circular equatorial orbits as they are accreted by the black hole and, oftentimes, that there is no torque at the inner boundary of the disk. The observed photon number flux density $N_{\rm obs}$ is, then, given by the expression~\cite{Li05}
\be
N_{E_{\rm obs}} = N_0 \left( \frac{ E_{\rm obs} }{ {\rm keV} } \right)^2 \int \frac{1}{M^2} \frac{ \Upsilon r' dr'd\phi'}{\exp \left[ \frac{N_1}{g F^{1/4}} \left( \frac{ E_{\rm obs} }{ {\rm keV} } \right) \right] - 1},
\label{eq:Nobs}
\ee
where
\be
N_0 \equiv 0.07205 f_{\rm col}^{-4} \left( \frac{M}{M_\odot} \right)^2 \left( \frac{D}{{\rm kpc}}\right)^{-2}{\rm keV^{-1}~cm^{-2}~s^{-1}},
\ee
\be
N_1 \equiv 0.1331 f_{\rm col}^{-1} \left( \frac{\dot{M}}{10^{18}~{\rm g~s^{-1}}} \right)^{-1/4} \left( \frac{M}{M_\odot} \right)^{1/2}.
\ee
Here, $M$, $D$, $\dot{M}$ are the mass, distance, and mass accretion rate of the black hole, respectively, $E_{\rm obs}$ is the observed photon energy, $f_{\rm col}$ is the spectral hardening factor, $F(r')=\sigma T_{\rm eff}^4(r')$ is the emitted disk flux with an effective temperature $T_{\rm eff}$, $\sigma$ is the Stefan-Boltzmann constant, $g(r')$ is the photon redshift, $\Upsilon(r')$ is geometric factor which takes into account the radiation type (e.g., isotropic or limb-darkened), and $r'$ and $\phi'$ are polar coordinates in the image plane.

For a Kerr black hole, the observed spectrum of the number flux density was discussed in great detail by, e.g., Ref.~\cite{Li05}. For higher values of the spin of the black hole, the spectrum becomes harder, because the inner edge of the accretion disk extends to smaller radii, which causes the radiative efficiency and the disk temperature to increase. For larger values of the disk inclination $i$, the observed flux density at lower energies decreases, because it originates at larger disk radii, where the observed flux density is $\propto \cos i$, while the observed flux density increases at higher energies, because it originates at smaller radii, where the relativistic effects of boosting, beaming, and light bending become important. As expected, when the accretion rate increases, the observed flux density likewise increases, and the spectrum becomes harder for higher values of the spectral hardening factor. 

Reference~\cite{Li05} also studied the effects of a nonzero torque at the inner edge of the disk and of returning radiation. They showed that these effects can be compensated by an adjustment of the mass accretion rate and the spectral hardening factor in a disk model without returning radiation and a torque at the ISCO. The effect of limb darkening is particularly strong at high disk inclinations and leads to a lower observed flux density relative to the observed flux density when the emission is isotropic.

\begin{figure}[ht]
\begin{center}
\psfig{figure=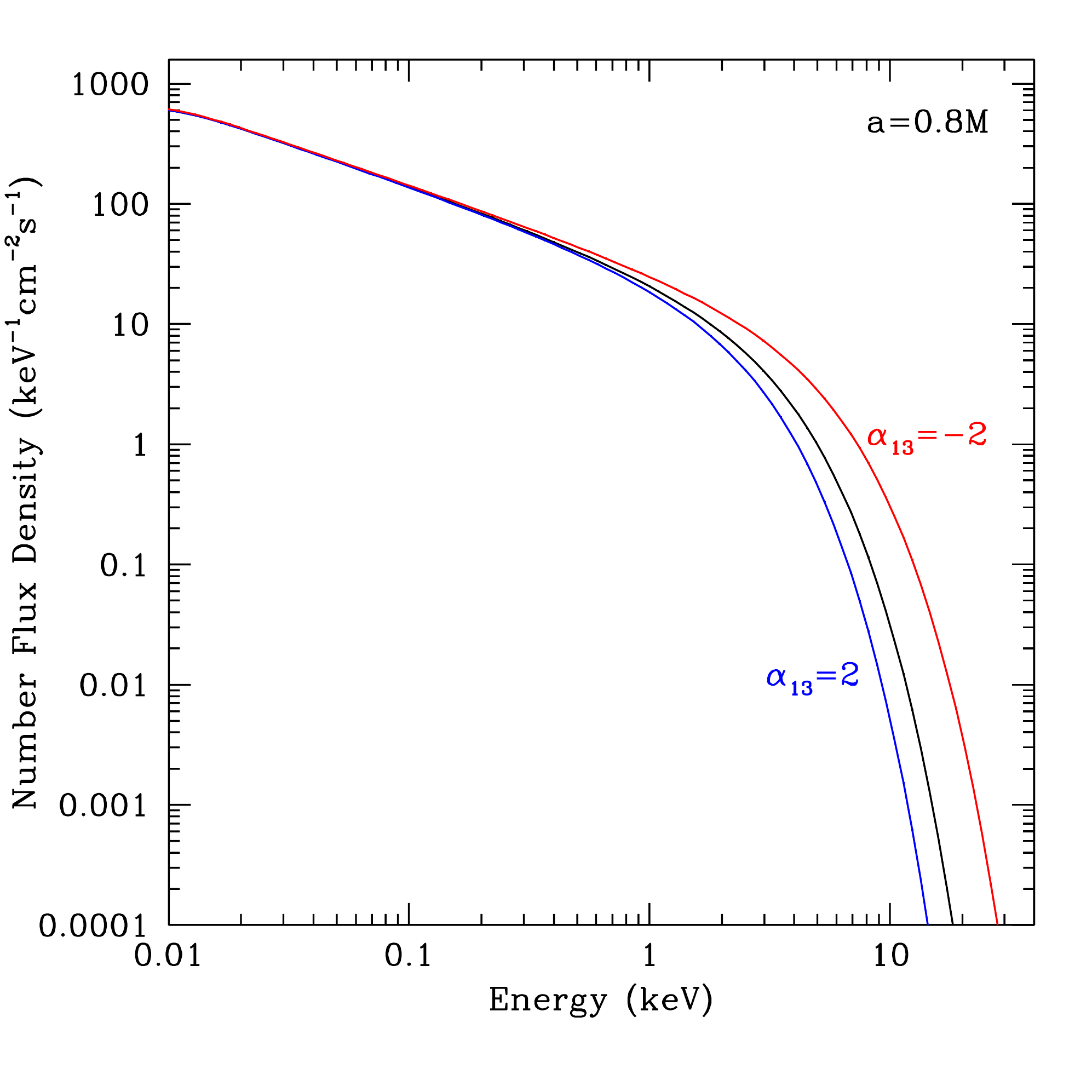,width=0.45\textwidth}
\psfig{figure=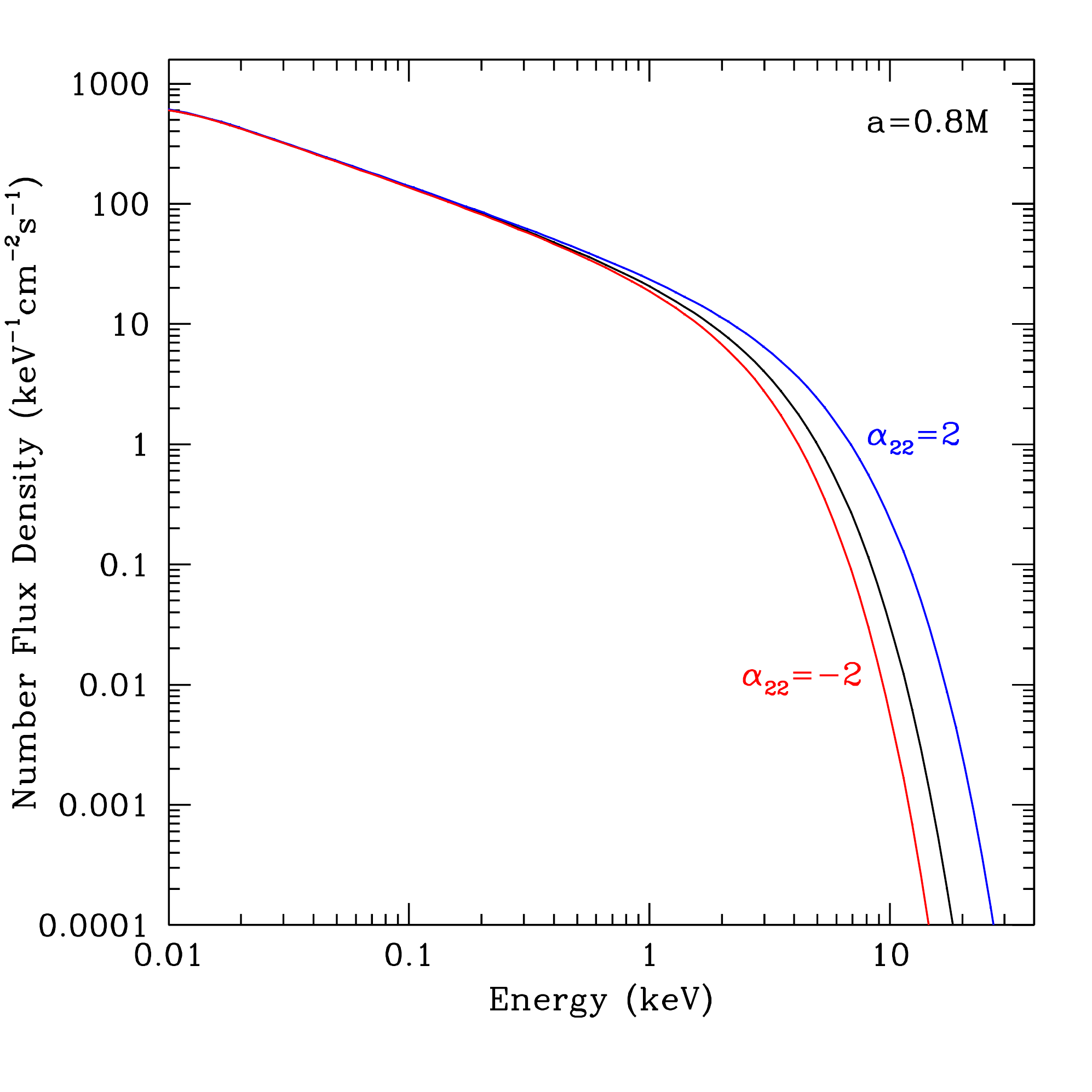,width=0.45\textwidth}
\end{center}
\caption{Observed thermal spectra from a geometrically thin accretion disk around black holes with spin $a=0.8r_g$, inclination $i=30^\circ$, mass $M=10M_\odot$, distance $D=10~{\rm kpc}$, mass accretion rate $\dot{M}=10^{19}~{\rm g~s^{-1}}$, and spectral hardening factor $f_{\rm col}=1.7$ for different values of the deviation parameters $\alpha_{13}$ (left) and $\alpha_{22}$ (right). The spectra become harder for decreasing values of the parameter $\alpha_{13}$ and increasing values of the parameter $\alpha_{22}$. For all spectra, an isotropic disk emission is assumed. Taken from Ref.~\cite{Xrayprobes}.}
\label{fig:spectra08}
\end{figure}

\begin{figure*}[ht]
\begin{center}
\psfig{figure=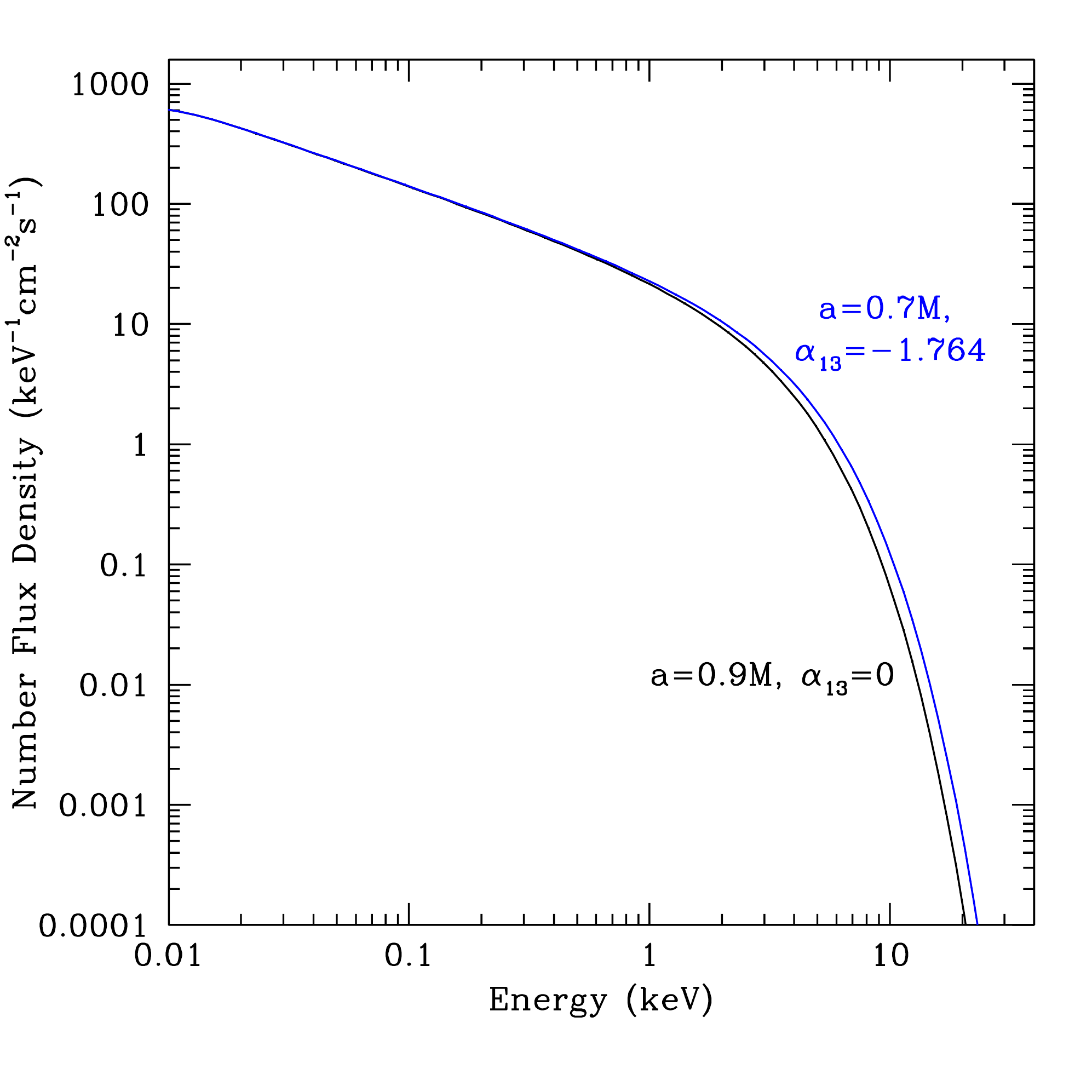,width=0.32\textwidth}
\psfig{figure=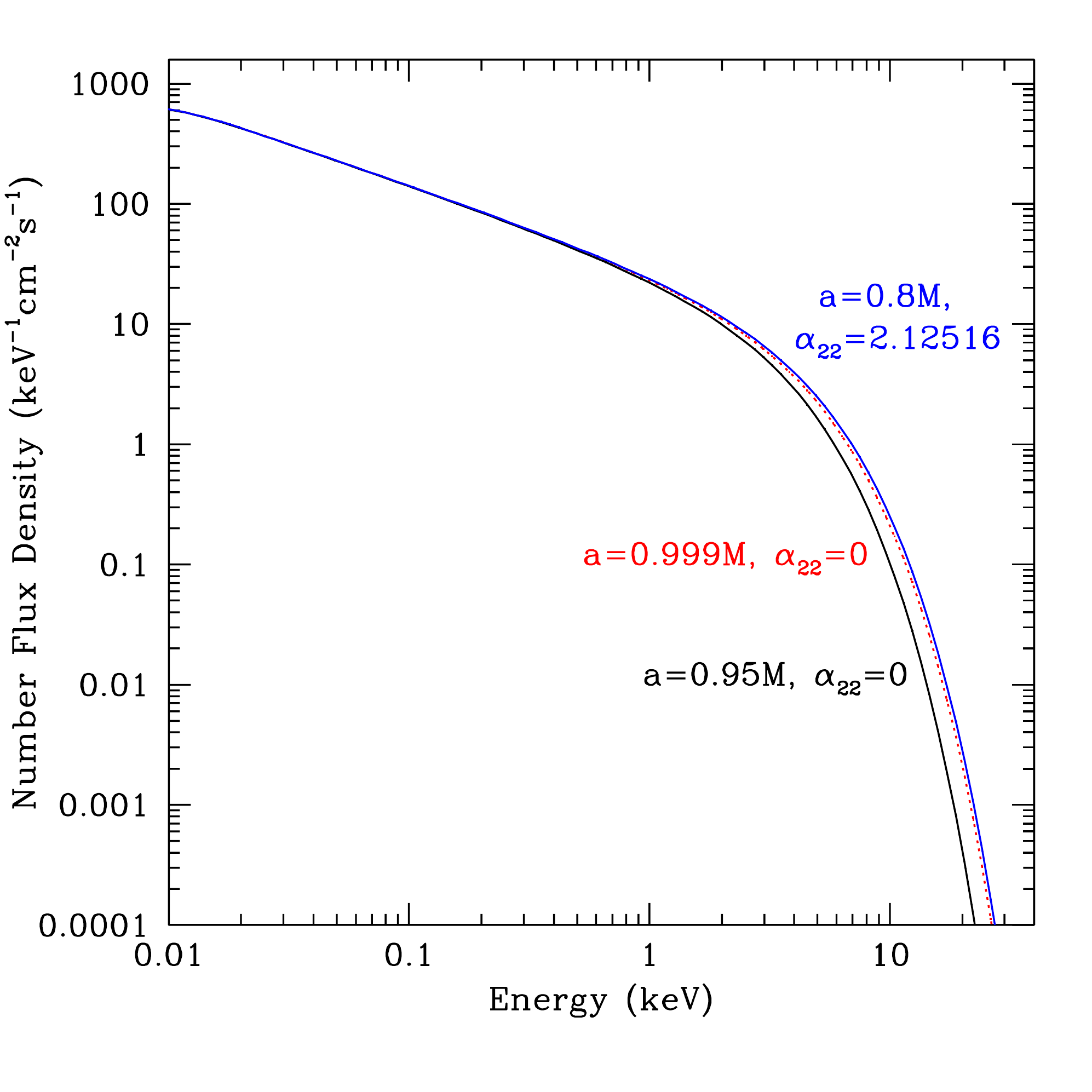,width=0.32\textwidth}
\psfig{figure=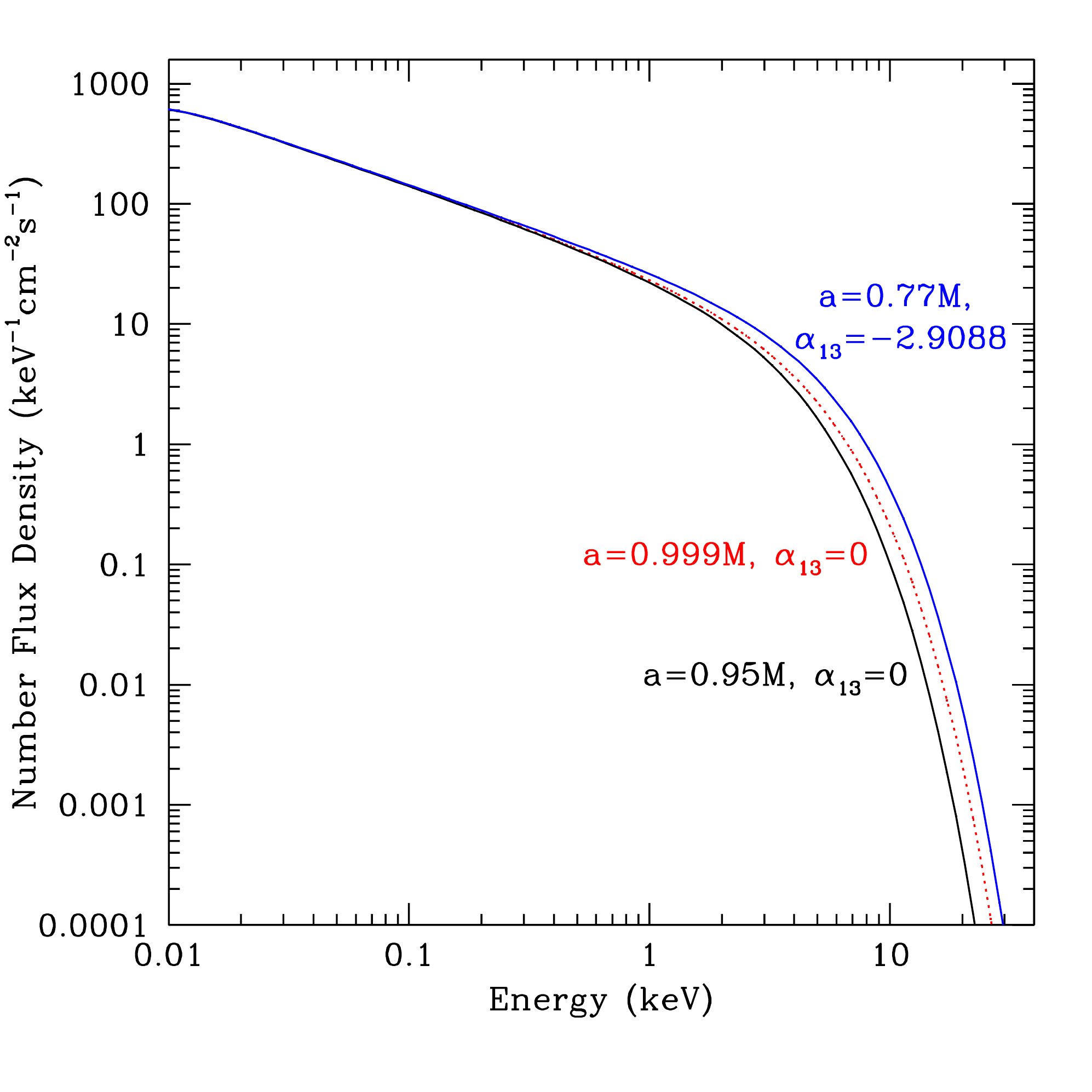,width=0.32\textwidth}
\end{center}
\caption{Observed spectra of the number flux density from a geometrically thin accretion disk around black holes with different values of the spin and deviation parameters $\alpha_{13}$ and $\alpha_{22}$ such that the ISCO coincides for the parameter combinations shown in each panel. The other parameters are the inclination $i=30^\circ$, the mass $M=10M_\odot$, the distance $D=10~{\rm kpc}$, the mass accretion rate $\dot{M}=10^{19}~{\rm g~s^{-1}}$, and the spectral hardening factor $f_{\rm col}=1.7$. The spectra differ significantly at the high-energy end, but some of the spectra of the Kerr-like black holes can still be confused with the spectra of a Kerr black hole with a different ISCO and spin. In the left panel, the spectrum of the Kerr-like black hole is very similar to the spectrum of a Kerr black hole with a spin $a\approx0.96r_g$ (not shown). In the center panel, the spectrum of the Kerr-like black hole can barely be mimicked by the spectrum of a Kerr black hole with a spin $a=0.999r_g$ (red dotted curve). In the right panel, however, the spectrum of the Kerr-like black hole cannot originate from a Kerr black hole even with a spin $a=0.999r_g$ (red dotted curve), because it does not extend to such high energies. Taken from Ref.~\cite{Xrayprobes}.}
\label{fig:spectra_isco}
\end{figure*}

Thermal continuum spectra for Kerr-like black holes described by the metric of Ref.~\cite{Jmetric} were analyzed by Ref.~\cite{Xrayprobes} (as a function of the parameters $\alpha_{13}$, $\alpha_{22}$, $\alpha_{52}$, and $\epsilon_3$) and by Ref.~\cite{Krawcz16} (as a function of the parameter $\beta$). The effect of the deviations from the Kerr metric on the observed spectra depend on the particular parameter. The spectra are strongly affected by both the parameters $\alpha_{13}$ and $\alpha_{22}$ in a manner that is similar to the effect of changing the spin of the black hole. Figure~\ref{fig:spectra08} shows the observed spectrum of the number flux density for a black hole with mass $M=10M_\odot$ and spin $a=0.8r_g$ for different values of the deviation parameters $\alpha_{13}$ and $\alpha_{22}$. For decreasing values of the parameter $\alpha_{13}$ and increasing values of the parameter $\alpha_{22}$, the number flux density extends to higher energies, while the low-energy part of these spectra is practically unaffected, because the deviation parameters predominantly affect the inner part of the disk. Changing the disk inclination if one or more of the deviation parameters are nonzero has qualitatively the same effect as in the case of a Kerr black hole.

Since the emitted flux depends only marginally on the parameters $\epsilon_3$ and $\alpha_{52}$, the effect of these parameters on the observed spectra is very small, at least in the case of isotropic emission. Both parameters primarily affect the peak flux density of the emitted radiation and the overall normalization of the observed flux density, while nonzero values of the parameter $\epsilon_3$ also cause a slight shift of the location of the ISCO (see Fig.~6 in Ref.~\cite{Jmetric}). Since the normalization of the flux density is used in practice to infer the mass accretion rate or, equivalently, the disk luminosity in units of the Eddington luminosity (see, e.g., Ref.~\cite{McClintock06}), the parameters $\epsilon_3$ and $\alpha_{52}$ cannot be obtained from the flux density normalization. Conversely, since their effect on the normalization is negligible, the mass accretion rate can be inferred robustly even if these two parameters are nonzero.

In the case of limb-darkened emission, the deviation parameters $\epsilon_3$, $\alpha_{13}$, and $\alpha_{22}$ have an effect on the observed spectra that is comparable in magnitude and analogous to the effect of limb darkening for Kerr black holes (c.f., Ref.~\cite{Li05}). As before, the parameter $\alpha_{52}$ has only a very minor effect on the observed spectra. Consequently, the observed flux density depends strongly on the deviation parameters $\alpha_{13}$ and $\alpha_{22}$ in the case of isotropic emission, while the observed flux density depends strongly on the deviation parameters $\epsilon_3$, $\alpha_{13}$, and $\alpha_{22}$ in the case of limb-darkened emission. In all cases, these parameters affect the observed spectra in a manner that is different from the effect of changing the mass accretion rate, which can, therefore, be measured independently of these parameters from the normalization of the flux density.

The question of whether the spin and the deviation parameters can be measured independently has to be analyzed more carefully, as in the case of iron line spectra (see Sec.~\ref{sec:ironlines}). For small to intermediate values of the spin, the observed flux density is very similar for Kerr and Kerr-like black holes with the same ISCO radius and the corresponding observed spectra are practically indistinguishable. For high values of the spin, however, the observed spectra differ significantly. Figure~\ref{fig:spectra_isco} shows the observed number flux density for Kerr black holes with spins $a=0.9r_g$ and $a=0.95r_g$ and for Kerr-like black holes with smaller spins but a nonzero values of the parameters $\alpha_{13}$ or $\alpha_{22}$ chosen such that the ISCO coincides with the ISCO of the respective Kerr black hole. The observed spectra differ strongly at the high-energy end~\cite{Xrayprobes}.

Reference~\cite{Krawcz16} found similar results regarding the spectral shapes and the similarity of different spectra for black holes with the same ISCO radius as a function of the parameter $\beta$ assuming isotropic disk emission. Implementing the metric of Ref.~\cite{Jmetric} as a function of the parameter $\beta$ into the algorithm of Ref.~\cite{Xrayprobes}, I find a similar dependence of such spectra on the parameter $\beta$ also in the case of limb-darkened emission. For either emission type, increasing values of the parameter $\beta$ lead to a softer spectrum.

Nonetheless, spectra of Kerr-like black holes with nonzero values of the deviation parameters are still very similar to the spectra of Kerr black holes with a different spin and, therefore, a different ISCO. However, if the spin of the Kerr black hole is near maximal, spectra of Kerr-like black holes can extend to such high energies that they cannot originate from a maximally spinning Kerr black hole. This illustrates an important difference between the continuum fitting and iron line methods: At high spins, thermal disk spectra already differ when the deviation parameters $\alpha_{13}$, $\alpha_{22}$, or $\beta$ are small, while iron line profiles differ only if the deviations or the emissivity index are sufficiently large. On the other hand, for sets of (high) spin values and deviation parameters $\alpha_{13}$, $\alpha_{22}$, and $\beta$ with different ISCO radii, thermal spectra can still be very similar, while iron lines cannot be confused, at least in principle.

These results imply that the deviation parameters $\alpha_{13}$, $\alpha_{22}$, and $\beta$ can be measured independently of the spin if the radius of the ISCO is $\approx r_g$ and the magnitude of the deviation is sufficiently large. This also means that the location of the ISCO can be measured robustly as long as the ISCO radius is not too close to the black hole corresponding to Kerr black holes with small to intermediate values of the spin. Since the limb darkening does not affect the high-energy end of the spectrum, nonzero values of the parameter $\epsilon_3$ can always be closely mimicked by a Kerr black hole with a different spin.

Thermal continuum spectra originating from Novikov-Thorne accretion disks around black holes or other compact objects described by other Kerr-like metrics were analyzed by Refs.~\cite{Torres02,Pun08,Harko09a,Harko09b,Harko10,BambiBarausse11,Krawcz12,BambiDiskCode,Moore15,Lin16,Krawcz16,Rannu16} (see, also, Ref.~\cite{Fede15}). The properties of such spectra are comparable and depend only weakly on the underlying metric. The strong correlation between the spin and the (lowest-order) deviation parameter in the metric of Ref.~\cite{JPmetric} likewise persists in this non-Kerr background~\cite{Bambi13}; see Ref.~\cite{BambiReview} for a review.

Using existing data, several authors have obtained constraints on the deviation parameters of different metrics. As expected, such constraints are much tighter for black holes with ISCO radii $\approx r_g$ (such as, e.g., Cyg X--1~\cite{Gou11,Gou14}) corresponding to near-maximal spin values if the Kerr metric is assumed, because the allowed parameters space of the deviations is usually much smaller. To date, all of these constraints are based on direct comparisons of either entire spectra or simply ISCO radii with the results inferred from the original X-ray data assuming a Kerr black hole~\cite{BambiBarausse11,Kong14,Bambi14a,Bambi14b,Xrayprobes,Krawcz16} and will be similar in magnitude for the parameters $\alpha_{13}$, $\alpha_{22}$, and $\beta$ (c.f., the discussion in Ref.~\cite{Xrayprobes}). A full data analysis of the X-ray data in a non-Kerr spacetime has not been performed yet.

\section{Variability}
\label{sec:variability}

Quasi-periodic variability is another observable phenomenon that could be used to test the no-hair theorem in the case of either stellar-mass and supermassive black holes. Sgr~A$^\ast$ is of particular interest, because its variability at near-infrared/radio frequencies is observable with GRAVITY and the EHT. At present, the exact mechanism behind the observed variability in black holes remains unclear and a number of different models have been suggested. Future observations as well as (likely) refined theoretical modelling will be required to distinguish between these models and to make an interpretation of variability unambiguous.

\subsection{Sgr~A$^\ast$}

Proposed models to explain the observed variability in Sgr~A$^\ast$ include the sudden heating of hot electrons in a jet~\cite{Markoff01}, compact flaring structures (``hot spots'') on nearly circular orbits in the accretion flow around Sgr~A$^\ast$~\cite{Bro06a,Bro06b,Hamaus09} (c.f., Refs.~\cite{StellaVietri98,Abramowicz01,Abramowicz03}), the ejection of a plasma blob out of the accretion flow~\cite{vanderLan66,YusefZadehQPO06}, magnetohydrodynamic turbulence along with density fluctuations~\cite{Goldston05,Chan06,Ball16} and particle accelerations due to Rossby wave instabilities~\cite{TaggerMelia06,Falanga07} (c.f., Ref.~\cite{TaggerVarniere06}), and red noise~\cite{Do09}. Infalling material such as the gas cloud G2~\cite{Gillessen12,Pfuhl15} could also lead to a substantial flux increase over several months~\cite{Schartmann12}.

For a Kerr black hole, a measurement of the orbital period of a hot spot can be used to infer the spin of Sgr~A$^\ast$ (as well as its mass) and several authors have argued that Sgr~A$^\ast$ must be rotating based on observed rapid periodicities~\cite{Genzeletal03,AschenbachQPO04,BelangerQPO06,MeyerQPO06,Trippe07,Kato10,Dokuchaev15}. On the other hand, Rossby wave instabilities may naturally produce periodicities on the order of tens of minutes even if Sgr~A$^\ast$ is not spinning~\cite{TaggerMelia06,Falanga07}.

Deeper insight into the structure of such flares is expected to be gained by observations with instruments such as GRAVITY~\cite{GRAVITY} and with the EHT~\cite{Doele09a,Doele09b,Fish09}. References~\cite{VincentGRAVITY11,VincentGRAVITY14} simulated GRAVITY observations of such flares in different models and showed that moving and non-moving flares located at the ISCO radius can be distinguished even for faint flares with a K-band magnitude of 15 and that flares originating from a blob ejected from the accretion flow can be distinguished from other flare models if the blob is ejected at an inclination larger than $\sim45^\circ$ and the flare has a duration of $\gtrsim1.5~{\rm h}$ and a K-band magnitude roughly between 14 and 15.

References~\cite{Bro06a,Bro06b} designed a 3D hot spot model with a Gaussian density profile of an overdensity of non-thermal electrons in the accretion flow with an extent of a few gravitational radii. The EHT is expected to be able to detect such flares and their orbital periods via closure phase/closure amplitude analysis~\cite{Doelehotspot09} and via polarization measurements~\cite{FishHotspot09}. Reference~\cite{Johnsonhotspot14} estimated that the EHT can make such detections with a precision of $\sim5~{\rm \mu as}$ on timescales of minutes, which is comparable to the anticipated precision of GRAVITY for similar observations~\cite{GRAVITY,VincentGRAVITY11}. Reference~\cite{JohnsonAxis15} analyzed the lagged covariance between interferometric baselines of similar lengths but slightly different orientations and demonstrated that the peak in the lagged covariance indicates the direction and angular velocity of the accretion flow, thus enabling the EHT to measure these quantities.

Combined observations of several hot spots at different radii with GRAVITY or the EHT could be used as tracers of the spacetime potentially revealing the spin of Sgr~A$^\ast$ as well as quadrupolar deviations from the Kerr metric. Reference~\cite{PaperIII} pointed out that measurements of the orbital period of hot spots should be able to measure the spin of Sgr~A$^\ast$ in that model even if the no-hair theorem is violated, because the Keplerian frequency of a given hot spot at a fixed radius depends only weakly on deviations from the Kerr metric.

Reference~\cite{BambiHotspot1} considered a 2D hotspot with a Gaussian density profile located in the equatorial plane of the compact object in the metric of Ref.~\cite{JPmetric} assuming monochromatic emission. For increasing values of the deviation parameter, the width of the light curve decreases and the frequency of the hot spot increases, which is caused primarily by the corresponding decrease of the ISCO radius. For hot spots orbiting at the same ISCO radius around compact objects with different sets of values of the spin and deviation parameter that correspond to that radius, there is a slight phase shift between the primary and secondary curves in the spectrogram potentially allowing these signals to be distinguished if the ISCO can be determined independently~\cite{BambiHotspot1}. Reference~\cite{BambiHotspot2} considered a similar model, where the hotspot is located at a fixed (small) height above or below the equatorial plane and found slight changes of the brightness of the hot spot depending on its position.

\subsection{X-ray variability}
\label{subsec:QPOs}

X-ray quasiperiodic oscillations (QPOs) have been observed in both stellar-mass black holes~(see, e.g., Refs.~\cite{MR06,Belloni12}) and AGN~\cite{Gierlinski08,Reis12b}. QPOs can be divided into two general classes, high-frequency QPOs (roughly 40–450~Hz) and low-frequency QPOs (roughly 0.1–30~Hz), and their observed frequencies sometimes fall into ratios of small integers; see Ref.~\cite{MR06} for a detailed discussion. Suggested QPO models include the diskoseismology model~\cite{Peres97,Silbergleit01,Wagoner08}), the relativistic precession model~\cite{StellaVietri98}, the epicyclic resonance model~\cite{Abramowicz01,Kluzniak01,Abramowicz03}, as well as others (e.g., \cite{Rezzolla03,Ingram11}). Based on the observed (high-frequency) QPOs, constraints on the masses and spins of certain stellar-mass black holes (assuming a Kerr black hole) in specific models have been obtained by Refs.~\cite{Wagoner99,Wagoner01,Abramowicz01,Rezzolla03,Torok05a,Torok05b,Wagoner12,Motta14a,Motta14b,Stuchlik16}.

\begin{figure*}[ht]
\begin{center}
\psfig{figure=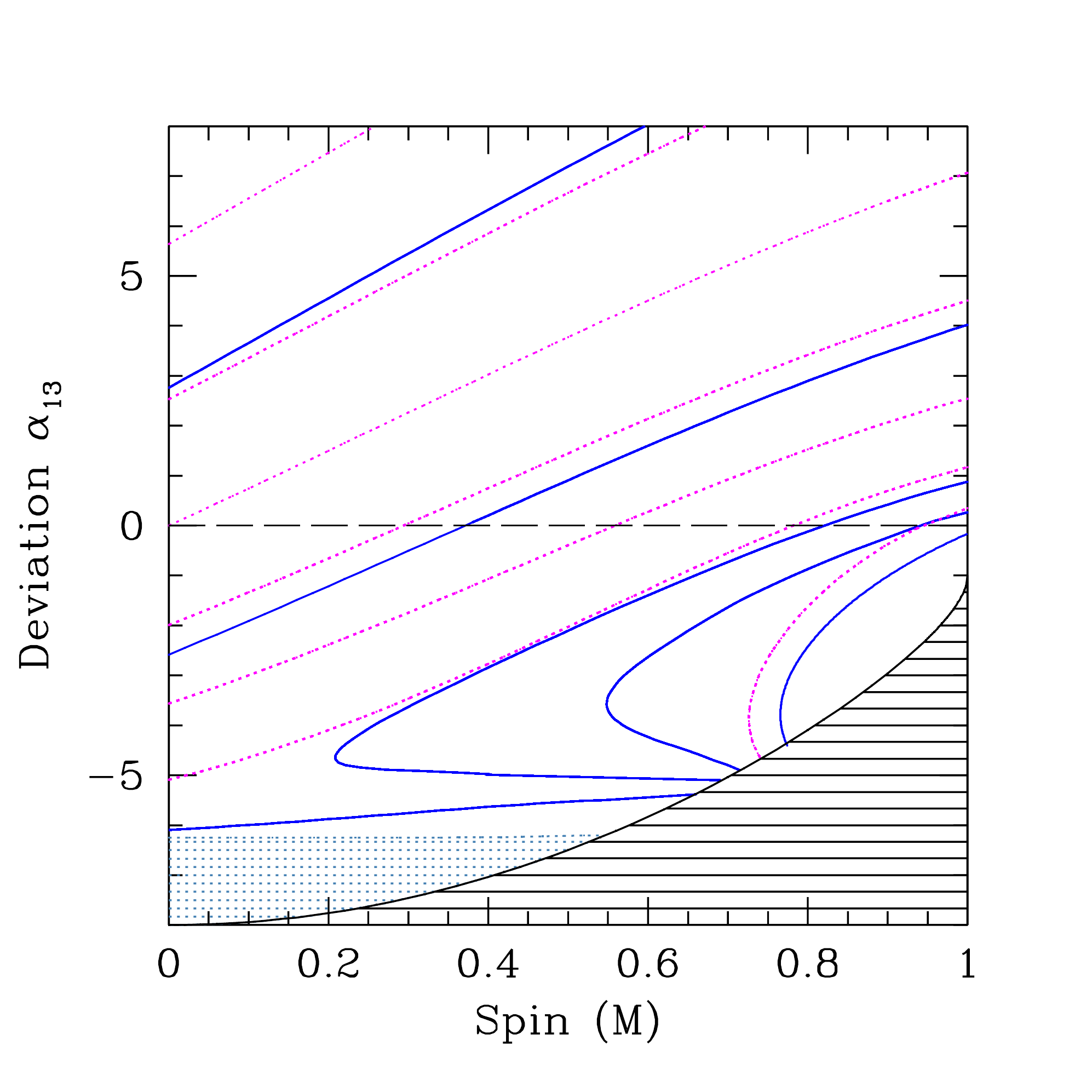,width=0.4\textwidth}
\psfig{figure=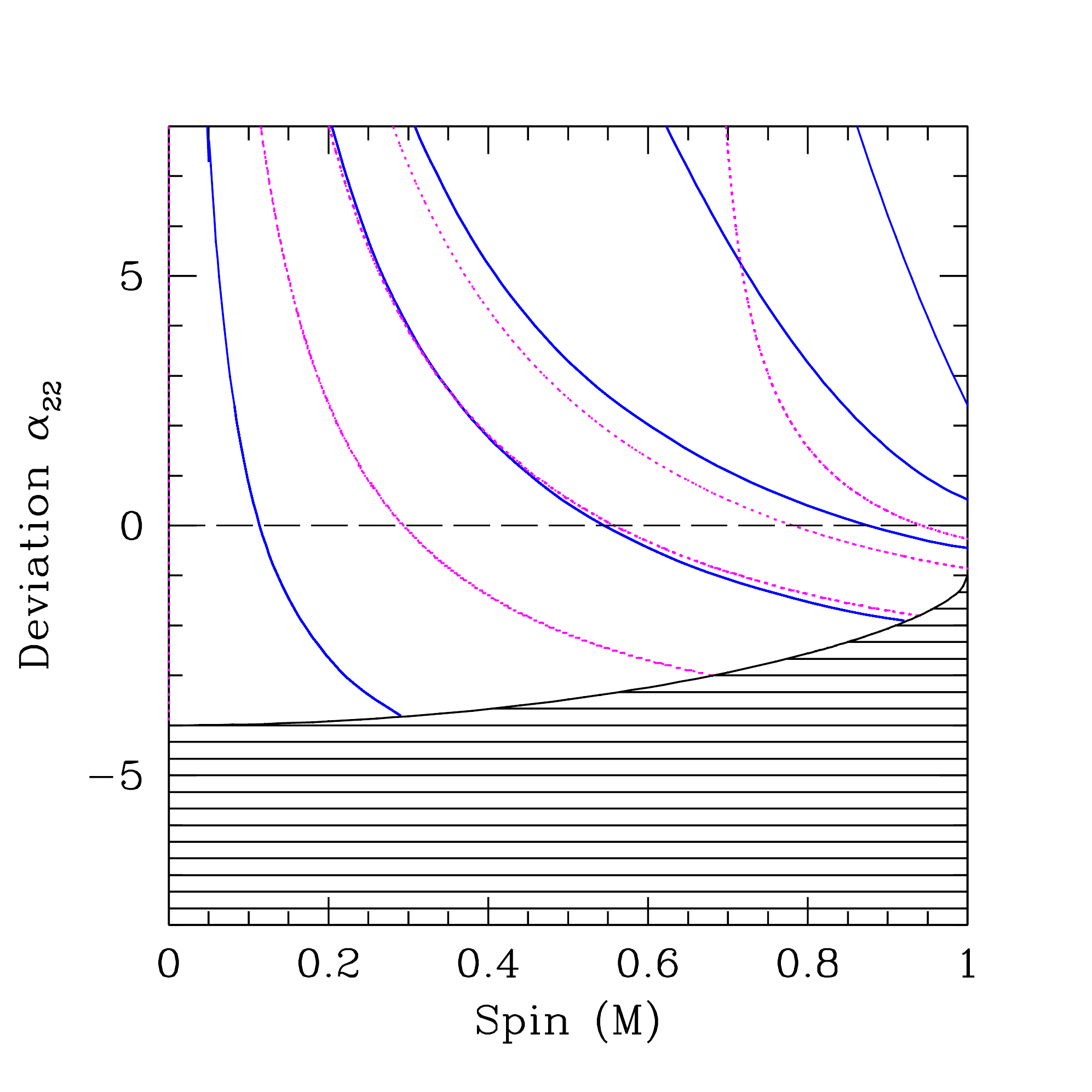,width=0.4\textwidth}
\end{center}
\caption{Contours of constant Keplerian frequency for a $3:1$ resonance between the Keplerian and radial epicyclic frequencies for a $10M_\odot$ black hole as a function of the spin and the deviation parameters $\alpha_{13}$ (left) and $\alpha_{22}$ (right). In the left panel, frequency contours are shown as solid blue curves with frequencies (top to bottom) $100~{\rm Hz}$, $150~{\rm Hz}$, $250~{\rm Hz}$, $450~{\rm Hz}$, $600~{\rm Hz}$, and $900~{\rm Hz}$. For reference, contours of constant ISCO radius are shown as dotted magenta lines with radii (top to bottom) $8r_g$, $7r_g$, $\ldots$, $2r_g$. In the right panel, frequency contours are shown as solid blue curves with frequencies (left to right) $200~{\rm Hz}$, $300~{\rm Hz}$, $500~{\rm Hz}$, $1000~{\rm Hz}$, and $1500~{\rm Hz}$. Contours of constant ISCO radius are shown as dotted magenta lines with radii (left to right) $6r_g$, $5r_g$, $\ldots$, $2r_g$. In both cases, the contours of constant resonance frequency are mostly aligned with the contours of constant ISCO radius except for large values of the parameter $\alpha_{22}$. The black shaded region marks the excluded part of the parameter space. Taken from Ref.~\cite{Xrayprobes}.}
\label{fig:res}
\end{figure*}

Reference~\cite{Xrayprobes} investigated the dependence of QPO frequencies on the spin and the deviation parameters $\alpha_{13}$, $\alpha_{22}$, $\alpha_{52}$, and $\epsilon_3$ of the metric of Ref.~\cite{Jmetric} in the diskoseismology model and the epicyclic resonance model. In the diskoseismology model, QPOs can arise as so-called gravity modes ($g$-modes~\cite{Peres97}) and corrugation modes ($c$-modes~\cite{Silbergleit01}). In the resonance model, QPOs can be identified as resonances between the dynamical frequencies~\cite{Abramowicz01,Kluzniak01,Abramowicz03}. 

Expressions for the Keplerian as well as the radial and vertical epicyclic frequencies for particles on circular equatorial orbits can be found in Ref.~\cite{Jmetric}. The fundamental $g$-mode occurs at the radius where the radial epicyclic frequency reaches its maximum and the fundamental $c$-mode corresponds to the Lense-Thirring frequency (i.e., the difference between the Keplerian and vertical epicyclic frequencies) at the ISCO \cite{Peres97,Silbergleit01}. A parametric resonance between the dynamical frequencies usually occurs at a different radius, which is determined directly by the frequency ratio of the resonance~\cite{Abramowicz01,Kluzniak01,Abramowicz03}.

In the diskoseismology model, contours of the fundamental $g$-mode as a function of the spin and one deviation parameter are mostly aligned with the corresponding ISCO contours except for very high frequencies, while the contours of the fundamental $c$-modes in the same parameter space are less aligned with the corresponding ISCO contours, especially for very low frequencies in the case of deviations described by the parameter $\alpha_{13}$. In the epicyclic resonance model, frequency contours in this parameter space are likewise mostly aligned with the corresponding contours of constant ISCO radius. Figure~\ref{fig:res} shows contours of constant Keplerian frequency in a $3:1$ resonance with the radial epicyclic frequency together with contours of constant ISCO radius.

Calculating the epicyclic frequencies as a function of the parameter $\beta$ [c.f., Eqs.~(83), (96), (97) in Ref.~\cite{Jmetric}] and computing the $g$-, $c$-, and $3:1$ resonance modes as discussed in, e.g., Sec.~VI of Ref.~\cite{Xrayprobes}, I find that these QPO modes likewise depend strongly on the parameter $\beta$. The dependences of the QPO modes on the spin and the deviation parameters $\alpha_{13}$, $\alpha_{22}$, and $\beta$ as discussed above are similar to the ones in other Kerr-like metrics~\cite{PaperIII,PaperIV,BambiQPO,Bambi15,Maselli15} (see, also, Ref.~\cite{Fede16}). However, since at the lowest order in the deviation functions the metric of Ref.~\cite{Jmetric} contains five deviation parameters instead of only one or two, the departure of the frequency contours from the contours of constant ISCO radius are generally larger. Since the dynamical frequencies and the location of the ISCO depend only marginally on the deviation parameter $\epsilon_3$ and not at all on the deviation parameter $\alpha_{52}$ (apart from the radial epicyclic frequency, which is affected only slightly), their effect on the $g$-, $c$-, and resonance modes are small~\cite{Xrayprobes}.

\section{X-ray polarization}
\label{sec:polarization}

Polarized X-ray emission could likewise be used as a probe of the spacetimes of stellar-mass black holes with observations by future missions such as the X-ray Imaging Polarimetry Explorer (XIPE;~\cite{Soffitta13}) or the Polarization Spectroscopic Telescope Array (PolSTAR;~\cite{Krawcz16a}). Signatures of such polarized emission for Kerr black holes have been analyzed by Refs.~\cite{Schnittman09,Schnittman10,Schnittman16} finding that typical polarization fractions are of the order of a few degrees and that the polarization fractions and orientations depend on the spin of the black hole.

\begin{figure*}[ht]
\begin{center}
\psfig{figure=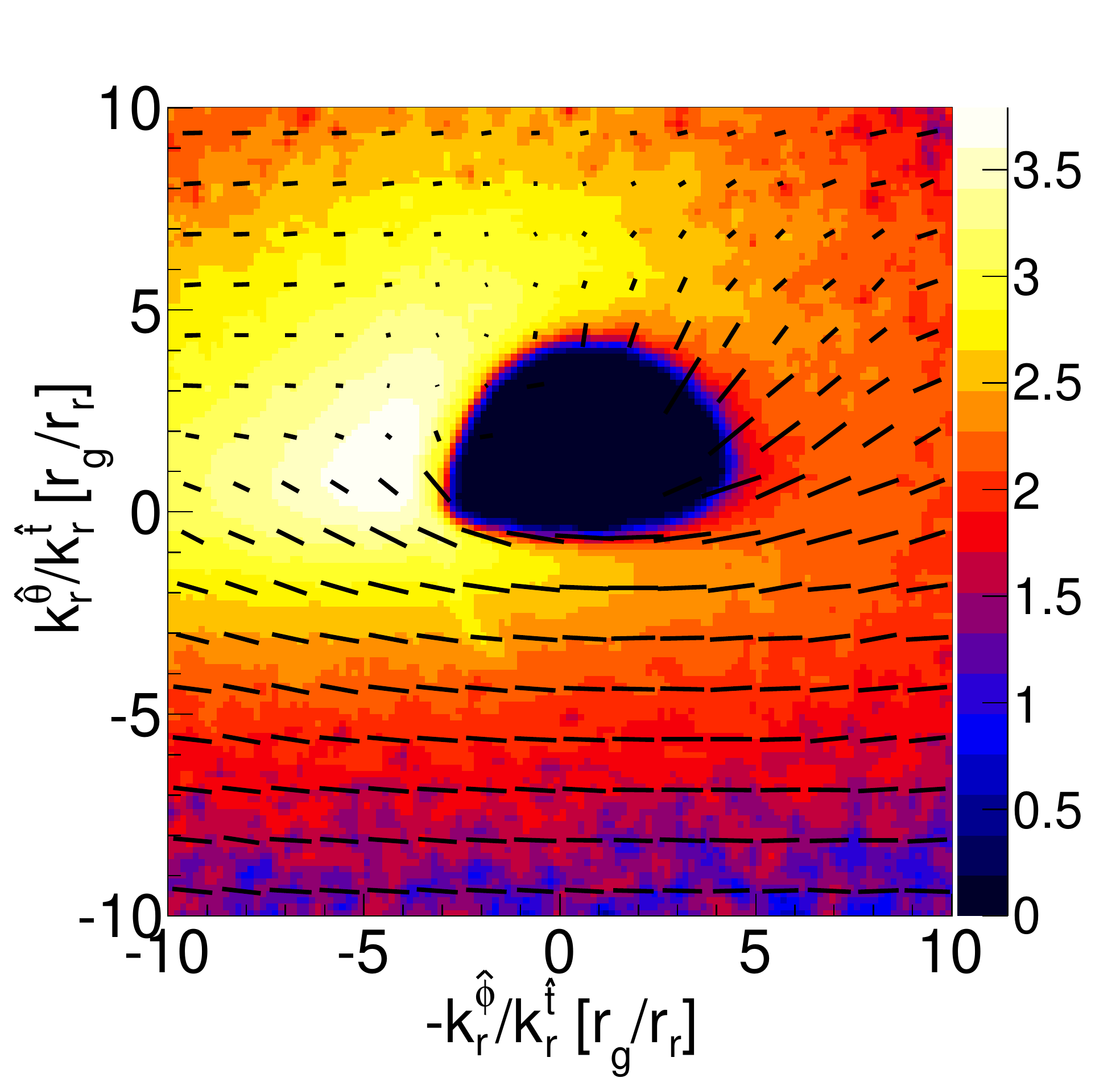,width=0.37\textwidth}
\psfig{figure=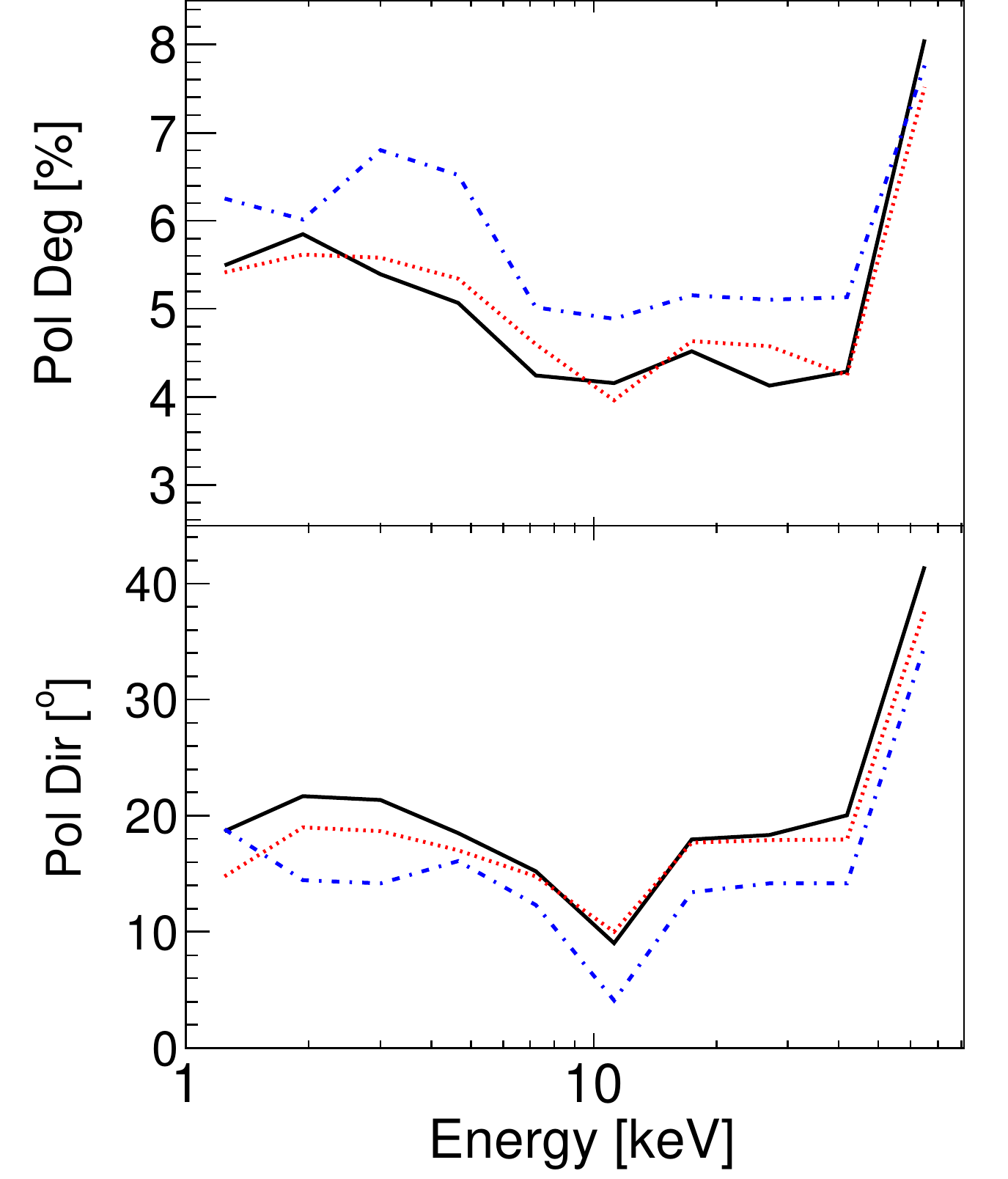,width=0.32\textwidth}
\end{center}
\caption{The left panel shows the intensity-polarization map of a Kerr-like black hole with values of the spin $a=0.5r_g$ and deviation parameter $\beta=0.69$. The length and direction of the black dashed lines indicate the polarization orientation and fraction, respectively. The right panel shows spectra of the (top) polarization fractions and (bottom) polarization orientations for the same Kerr-like black hole (blue dash-dotted curve), as well as for a Kerr black hole with a value of the spin $a=0.9r_g$ (black solid curve) and a Kerr-like compact object described by the metric of Ref.~\cite{JPmetric} (red dotted curve). In all three cases, the ISCO is located at the same coordinate radius. While the spectra are similar, the deviations described by the parameter $\beta$ are slightly larger than the deviations in the metric of Ref.~\cite{JPmetric}. Taken from Ref.~\cite{Krawcz16}.}
\label{fig:polarization}
\end{figure*}

References~\cite{Krawcz12,BambiPol,Krawcz16} analyzed (isotropic) polarized emission originating from the accretion disks around Kerr-like black holes in different metrics and showed that the polarization fractions and orientations depend on the respective deviation parameter(s) from the Kerr metric. Figure~\ref{fig:polarization} shows the intensity-polarization map of a Kerr-like black hole with values of the spin $a=0.5r_g$ and deviation parameter $\beta=0.69$ in the metric of Ref.~\cite{Jmetric} together with a comparison of the corresponding spectra of the polarization fraction and polarization orientation with those of a Kerr black hole with spin $a=0.9r_g$ and a Kerr-like compact object decribed by the metric of Ref.~\cite{JPmetric}, all three of which have the same ISCO radius. The spectra are similar, although the deviations described by the parameter $\beta$ are slightly larger than the deviations in the metric of Ref.~\cite{JPmetric} (c.f., Ref.~\cite{Krawcz12}).

\section{Discussion}
\label{sec:conclusions}

Thus far, general relativity has passed all tests. However, these have focused almost exclusively on the weak-field regime~\cite{Will14}, while the strong-field regime still remains practically untested~\cite{Psaltis08}. In the next few years and coming decades, a number of different experiments across the electromagnetic spectrum will probe the latter regime around supermassive and stellar-mass black holes with unprecedented precision, opening the door for a second look at the validity of the no-hair theorem and general relativity.

Table~\ref{tab:dependencies} summarizes the dependence of the strong-field observables discussed in this article on the lowest-order deviation parameters of the metric of Ref.~\cite{Jmetric} to the extent that they have been investigated. The parameters $\alpha_{13}$, $\alpha_{22}$, and $\beta$ typically have a strong effect on all of these observables, while the parameters $\alpha_{52}$ and $\epsilon_3$ only affect the X-ray observables in certain cases. Weak-field probes such as the observations of stars and pulsars orbiting around Sgr~A$^*$ aim to measure the quadrupole moment of the black hole directly, which, in turn, would place constraints on the deviation parameters (see, e.g., Refs.~\cite{GB06,Suvorov15}).

\begin{table}[ht]
\begin{center}
\footnotesize
\begin{tabular}{lllllll}
\multicolumn{7}{c}{}\\
Observable   & $\alpha_{13}$ &~$\alpha_{22}$ & $\alpha_{52}$ & $\epsilon_3$ & $\beta$ & References \\
\hline
ISCO location  & Strong      & Strong        & None          & Moderate     & Strong  & \cite{J13rings,JohannsenReview} \\
Shadow size  & Strong        & Weak          & None          & None         & Strong  & \cite{J13rings,JohannsenReview} \\
Shadow displacement & Strong    & Strong        & None          & None      & Strong  & \cite{J13rings,JohannsenReview} \\
Shadow asymmetry & Strong    & Strong        & None          & None         & Strong  & \cite{J13rings,JohannsenReview} \\
Thermal spectrum (isotropic) & Strong   & Strong   & Weak   & Weak          & Strong  & \cite{Xrayprobes,Krawcz16} \\
Thermal spectrum (limb darkened) & Strong   & Strong   & Weak   & Strong    & Strong  & \cite{Xrayprobes,thiswork} \\
Iron line (isotropic) & Strong   & Strong   & Weak   & Weak          &  Strong        & \cite{Xrayprobes,thiswork} \\
Iron line (limb darkened) & Strong   & Strong   & Moderate   & Moderate     & Strong  & \cite{Xrayprobes,thiswork} \\
QPO (diskoseismology) & Strong   & Strong   & Weak   & Weak   &  Strong               & \cite{Xrayprobes,thiswork} \\
QPO (resonance)       & Strong   & Strong   & Weak   & Weak   &  Strong               & \cite{Xrayprobes,thiswork} \\
X-ray polarization (isotropic)  &   &   &   &   &   Strong                            & \cite{Krawcz16} \\
\hline
\end{tabular}
\caption{Dependence of different strong-field observables on the (lowest-order) deviation parameters of the Kerr-like metric of Ref.~\cite{Jmetric}.}
\label{tab:dependencies}
\end{center}
\end{table}

Ideally, the results of tests of the no-hair theorem with different methods would be combined to provide the strongest constraints on potential deviations from the Kerr metric. This could be achieved in several scenarios which include the targets (i) Sgr~A$^*$ with near-infrared, pulsar-timing, and EHT observations, (ii) stellar-mass black holes with two or more of the X-ray observables, (iii) supermassive black holes besides Sgr~A$^*$ (and probably M87) with iron line and QPO observations, and (iv) pulsar black-hole binaries (if discovered) with pulsar-timing and X-ray observations.

The EHT has high promise to provide strong constraints on such deviations, in particular if a nearly circular shadow is observed in the case of Sgr~A$^*$~\cite{JohannsenPRL}. This is also the case for pulsar-timing observations if a suitable pulsar black-hole binary can be found~\cite{Liu12,Liu14}, as well as for long-term near-infrared observations of stars orbiting around Sgr~A$^*$ with instruments such as GRAVITY~\cite{GRAVITY}. Observations on event horizon scales of supermassive black holes apart from Sgr~A$^*$ and M87 with the EHT would require future VLBI stations in space~\cite{Hirabayashi98,Zhakarov05,Kardashev09,Wild09}. 

All of the X-ray observables are affected by a strong correlation between pairs of the spin and the deviation parameters which correspond to the same ISCO radius, although this correlation can be significantly reduced if the ISCO radius is $\approx r_g$ (corresponding to rapidly-spinning Kerr black holes) or in the case of low-frequency $c$-modes. While such a strong correlation was also found in initial accretion-flow studies with the EHT~\cite{Bro14}, it does not persist for EHT observations with larger arrays~\cite{Johannsenetal16}. For X-ray observables, high-spin sources will typically be optimal, because, in addition to the reduced correlation between the spin and the deviation parameters, the available parameter space is much smaller than it is for black holes with low to moderate spins~\cite{Xrayprobes,Krawcz16}.

Detections of fluorescent iron lines with a high signal-to-noise ratio as expected for future X-ray missions such as Athena+~\cite{Nandra13} will probably yield the strongest constraints in the X-ray spectrum. Future timing missions such as the Large Observatory For x-ray Timing (LOFT;~\cite{Feroci14}) should be able to resolve higher-order harmonics of QPOs observed in stellar-mass black holes which, in turn, would allow for a distinction between different QPO models. Observations of X-ray polarization will require future missions such as XIPE~\cite{Soffitta13} or PolSTAR~\cite{Krawcz16a}.

A different test of the no-hair theorem might be performed with the observation of black holes if space has more than three dimensions. In Randall-Sundrum-type braneworld gravity~\cite{RandallSundrum2}, black holes are unstable and evaporate into the extra dimension~\cite{Emparan03,Tanaka03}; see Ref.~\cite{Maartens04} for a review. This mechanism, then, allows for the size of the extra dimension to be constrained by the age of black holes~\cite{Psaltis07} and by the expected orbital evolution of black-hole binaries~\cite{Braneworld1,Braneworld2,McWilliams10,Simonetti11,Braneworld4}. Since the evaporation rate scales with the inverse cube of the black-hole mass, this effect is only relevant for stellar-mass black holes and completely negligible for supermassive black holes~\cite{Emparan03,Tanaka03}, thus not affecting tests of the no-hair theorem with observations of Sgr~A$^*$ or M87.

Likewise, black holes may also provide a proving ground of quantum effects that are large enough to be observable. The information paradox implies that the fundamental principles of relativity, quantum mechanics, and locality cannot be reconciled; see Ref.~\cite{Mathur09} for a review. One proposed solution to this conundrum is to introduce a modified notion of locality based on a subsystem structure of the physical system~\cite{Giddings12} within which a more fundamental Hamiltonian governs the information transfer (e.g.,  \cite{Giddings92}). Such a setup, then, leads to a different nature of black holes as encapsulated in the ``firewall''~\cite{AMPS13} or ``fuzzball''~\cite{Mathur08} scenarios. In the former, black holes exhibit an increased energy flux (e.g.,~\cite{Giddings12b}) and quantum metric fluctuations around them on scales of their horizon radii~\cite{Giddings14}. The latter effect could potentially disrupt an accretion flow near the horizon, distort the shadow of the black hole, and cause scintillation of light passing close to the black hole, as well as an alteration of gravitational wave emission from inspirals~\cite{Giddings14,Giddings16}.

Timing observations of the quasar OJ~287 could yield another test of the no-hair theorem within a timing model which is based on gravitational-wave emission of two inspiralling supermassive black holes. In this model, the smaller supermassive black holes passes through the accretion disk of the primary at the observed $\sim12~{\rm yr}$ periodicity producing significant outbursts. Such a test may be possible at the next outburst in 2023~\cite{Valtonen97,Valtonen08,Valtonen11,Valtonen16}.


This work was supported in part by Perimeter Institute for Theoretical Physics. Research at Perimeter Institute is supported through Industry Canada and by the Province of Ontario through the Ministry of Research \& Innovation.

\end{document}